\documentclass[fleqn,usenatbib]{mnras}

\usepackage{newtxtext,newtxmath}

\usepackage[T1]{fontenc}
\usepackage{ae,aecompl}

\usepackage{graphicx}	
\usepackage{amsmath}	
\usepackage{amssymb}	

\title[Sensitivity of HERA to One-point Statistics]{Sensitivity of the Hydrogen Epoch of Reionization Array and its Build-out Stages to One-point Statistics from Redshifted 21~cm Observations}

\author[P. Kittiwisit et al.]{Piyanat Kittiwisit$^{1}$\thanks{Contact e-mail: \href{mailto:piyanat.kittiwisit@asu.edu}{piyanat.kittiwisit@asu.edu}},
Judd D. Bowman$^{1}$,
Daniel C. Jacobs$^{1}$,
\newauthor{Adam P. Beardsley$^{1}$,
and Nithyanandan Thyagarajan$^{1,2}$\thanks{Nithyanandan Thyagarajan is a Jansky Fellow of the National Radio
Astronomy Observatory.}}
\\
$^{1}$School of Earth and Space Exploration, Arizona State University, AZ, USA\\
$^{2}$National Radio Astronomy Observatory, 1003 Lopezville Road, Socorro, NM
87801, USA}
\date{Accepted XXX. Received YYY; in original form ZZZ}

\pubyear{2017}

\begin{document}
\label{firstpage}
\pagerange{\pageref{firstpage}--\pageref{lastpage}}
\maketitle

\begin{abstract}
We present a baseline sensitivity analysis of the Hydrogen Epoch of Reionization Array (HERA) and its build-out stages to one-point statistics (variance, skewness, and kurtosis) of redshifted 21~cm intensity fluctuation from the Epoch of Reionization (EoR) based on realistic mock observations. By developing a full-sky 21 cm lightcone model, taking into account the proper field of view and frequency bandwidth, utilising a realistic measurement scheme, and assuming perfect foreground removal, we show that HERA will be able to recover statistics of the sky model with high sensitivity by averaging over measurements from multiple fields. All build-out stages will be able to detect variance, while skewness and kurtosis should be detectable for HERA128 and larger. We identify sample variance as the limiting constraint of the variance measurement while skewness and kurtosis measurements will be primarily limited by thermal noise. The sensitivity can be improved by performing frequency binning and windowing. In addition, we find that strong sample variance fluctuation in the kurtosis measured from an individual field of observation indicates the present of outlying cold or hot regions in the underlying fluctuations, a feature that can potentially be used as an EoR bubble indicator.

\vspace{2ex}
\noindent \emph{This manuscript is altered from the originally published paper \citep{2018MNRAS.474.4487K} to reflect corrections in the erratum \citep{2018MNRAS.477..864K}}.
\end{abstract}




\section{Introduction}
\label{sec:intro}

Considerable effort is underway to constrain the Epoch of Reionization (EoR), an era when radiation from the first stars and galaxies transformed gas in the intergalactic medium (IGM) from neutral to ionised. Observations of the Lyman-$\alpha$ forest in high-redshift quasars have set a limit to the end of reionization of $z\sim6.5$ \citep{2006AJ....132..117F}, and measurements of the cosmic microwave background (CMB) optical depth by the Planck experiment have indicated that reionization is still progressing at $z\sim8.8$ \citep{2016A&A...594A..13P}.

The 21~cm emission from the hyperfine transition in the ground state of neutral hydrogen is arguably the most direct probe to detect the EoR \citep{1972A&A....20..189S,1990MNRAS.247..510S,1997ApJ...475..429M,2000ApJ...528..597T,2002ApJ...572L.123I}. Full-sky observations of 21~cm spectra, redshifted to metre-wave, will produce tomographic maps of neutral hydrogen throughout the reionization era and beyond, allowing the study of the evolution of this structure and its implication for the underlying ionizing sources.

Many telescopes have been built to conduct experiments aiming to map this signal. These include the MWA \citep[Murchison Widefield Array;][]{2013PASA...30....7T,2013PASA...30...31B}, LOFAR \citep[Low Frequency Array;][]{2013A&A...556A...2V} and PAPER \citep[Donald C. Backer Precision Array for Probing the Epoch of Reionization;][]{2010AJ....139.1468P}. These telescopes utilise many compact antennas to yield tens-of-degrees fields of view and sub-degree to arcminute angular resolutions. They are also capable of observing with narrow ($\lesssim$10 kHz) spectral channel and simultaneously cover wide ($\sim100$ MHz) frequency bandwidth. These characteristics are ideal for EoR tomographic mapping but also give rise to widefield beam chromaticity and strong sidelobe interferences that complicate mitigation of bright astrophysical foregrounds. 

Due to sensitivity limitations of the present experiments, much attention has been focused on the statistical analysis of redshifted 21~cm observations, in particular with the power spectrum measurements. Upper limits from current observations have recently been released \citep{2013MNRAS.433..639P,2014ApJ...788..106P,2014PhRvD..89b3002D,2015PhRvD..91l3011D,2015ApJ...809...61A,2015ApJ...801...51J,2016ApJ...833..102B,2017ApJ...838...65P}, including robust characterisation of the foreground contamination and instrumental systematics in the power spectrum \citep{2012ApJ...752..137M,2013ApJ...770..156H,2013ApJ...769....5J,2014PhRvD..90b3019L,2015ApJ...804...14T,2015ApJ...807L..28T,2016ApJ...825....9T}. 

Lessons from the first generation of experiments have led to the design of HERA \citep[Hydrogen Epoch of Reionization Array;][]{2017PASP..129d5001D}, which is currently being built in the Karoo desert in South Africa. HERA will be a highly-packed, redundant-baseline array that is optimised for the power spectrum analysis while also retaining imaging performance on sub-degree scales. When completed, it will consist of 350, zenith-pointed, 14-metre dishes fed by dual-polarisation dipoles, with most dishes closely packed into a hexagon core approximately 300~meters in diameter and a small number of dishes spreading around the hexagon core to improve imaging performance. The SKA \citep[Square Kilometre Array;][]{2013ExA....36..235M}, which will be a multinational large scale radio observatory, is also scheduled to be built in the next decade. When completed, the SKA will be able to achieve high performance for both the power spectrum and direct imaging of the EoR. Statistical analysis of the EoR, however, will remain important for the next decade. 

As reionization progresses, ionised H\,{\scriptsize II} regions will form around groups of sources with high-energy UV radiation, causing the distribution of 21~cm intensity field to deviate from the nearly Gaussian underlying matter density field \citep{2006MNRAS.372..679M,2007ApJ...659..865L}, and the power spectrum alone will be insufficient to fully describe the signal. This limitation has motivated several theoretical studies on alternative statistics. One promising method is the measurement of the one-point probability distribution function (PDF) and higher-order one-point statistics (variance, skewness and kurtosis) of the 21~cm brightness temperature fluctuations. These statistics exhibit unique evolution throughout the reionization redshifts with distinct non-Gaussian features. For example, skewness and kurtosis are shown to sharply increase near the end of reionization when only isolated islands of 21~cm emission remain \citep{2007MNRAS.379.1647W,2009MNRAS.393.1449H,Shimabukuro:2015cpa,2016MNRAS.456.3011D}. Recent studies by \citet{Watkinson:2014jv,Watkinson:2015ce} and \citet{2015MNRAS.449.3202W} suggest that next-generation 21 cm arrays will be able to measure variance and skewness to distinguish different reionization models with high sensitivity.

In this work, we establish a detailed baseline expectation for 21~cm one-point statistics for HERA, focusing on expected thermal uncertainty and sample variance.  We consider realistic observing scenarios and investigate the performance of various phases of the HERA deployment with increasing numbers of antennas operating.  Our simulations  incorporate the effects of time-evolution of the signal and mapping of time-dependent redshift-space to frequency.

We describe our simulation in Section~\ref{sec:sim}. We provide more background information of the HERA instruments and observations in Section~\ref{sec:hera}. We develop our sky model in Section~\ref{sec:model} and present its statistics as a reference for this work in Section~\ref{sec:ref_stats}. We describe out mock observations in Section~\ref{sec:obs} and discuss the resulting measurements in Section~\ref{sec:results}. In Section~\ref{sec:bandwidth}, we look into a way to further improve sensitivity of the measurements through bandwidth averaging. We consider performances of the planned built-out HERA configurations in Section~\ref{sec:performance}. We introduce a potential detection method that takes advantage of sample variance on the kurtosis measurements to identify ionised regions in the observing field in Section~\ref{sec:kurtosis}. Finally, we conclude in Section~\ref{sec:conclusion}.

\section{Simulations}
\label{sec:sim}

We construct our HERA simulation pipeline based on existing 21~cm models and a simple approximation of the HERA instrument. We describe each component in the following subsections. All simulated images are noiseless. Thermal uncertainty is included in the analysis using analytic formulas from \citet{Watkinson:2014jv}. Sample variance contributes to the uncertainty of 21~cm one-point statistics due to the limited field of view and angular resolution of telescope. We perform Monte Carlo simulations to estimate sample variance. We ignore foreground contamination, postponing it to a future work.

\subsection{HERA}
\label{sec:hera}
HERA is a second-generation radio interferometer optimized for redshifted 21~cm power spectrum detection. Presently under construction, HERA uses large, 14-metre parabolic dishes as antenna elements with most dishes densely packed into hexagon shape to increase the sensitivity at the short baselines and aid with calibration. A number of outriggers are spread around the hexagon core to improve imaging performance. The construction of the telescope will be divided into 5 stages. As of 2017, the first stage with 19 dishes has been completed and commissioned. The second stage with 37 dishes is being constructed at the time of writing of this manuscript, and the third, fourth and the final fifth stages with 128, 240 and 350 dishes are scheduled to be constructed in 2017, 2018 and 2019. Observations with each build-out stage will be conducted subsequently following each construction period. For redshifted 21~cm observations, only the hexagon core will be used, yielding a maximum angular resolution of $\sim0.5$ deg from the full array. The telescope will have a $\sim9^{\circ}$ primary field of view and will operate between 100 and 200~MHz with a channel bandwidth of 100 kHz.

Pointed at the zenith at all time, HERA will observe in a drift-scan mode and records $\approx0.7$ hours of integration for each observed field on the sky per day. Only nighttime observations will be used for redshifted 21~cm science, resulting in 125 hours of integration per field per year. Assuming $\approx20$\% of observations are discarded due to poor weather conditions or radio-frequency interference, we expect an effective integration of 100~hours per year for any given region observed by HERA. As the array is located at -30$^{\circ}$ latitude, the Galactic Centre and anti-Centre pass almost overhead through the telescope beam. HERA will only compile 21~cm drift scans when the hight-Galactic latitudes are overhead, yielding a strip of a sky that spans approximately 180 degrees of Right Ascension nested between the Galactic plane.

However, simulating a full drift scans and making a single mosaic image from the observations are both very challenging. In this work, we approximate the drift scan by combining the analysis of multiple independent $\sim$9 degree fields that span the drift scan length. Less sky coverage of individual fields will yield measurements with higher sample variance, but we will show that statistics of the sky model can be recovered by averaging over multiple measurements from different fields, as well as deriving sample variance and thermal noise uncertainties of these measurements. For the rest of this paper, we will simply refer to these multiple-field measurements as drift scan measurements.

Additional details of the specifications of the HERA instruments and planned observations are in \citet{2016ApJ...826..181D} and \citet{2017PASP..129d5001D}.

\subsection{Sky Model}
\label{sec:model}
Our sky model consists of 21~cm brightness temperature intensity fluctuations ($\delta T_b(\nu)$), which are characterized by the 21~cm spin temperature ($T_S$),  the CMB temperature ($T_{\gamma}$), the fractional over-density of baryons $(1 +  \delta)$, and the gradient of the proper velocity along the line of sight ($\mathrm{d}v_{\parallel}/\mathrm{d}r_{\parallel}$),
\begin{gather}
    \begin{split}
        \delta T_b(\nu) \approx 9 x_{HI}&(1+\delta)(1+z)^{1/2} \\
        &\left[1 - \frac{T_{\gamma}}{T_S}\right]
         \left[\frac{H(z)/(1+z)}{\mathrm{d}v_{\parallel}/\mathrm{d}r_{\parallel}}
          \right]\ \mathrm{mK}.
    \end{split}
\end{gather}

Many current state-of-the-art 21~cm simulators use semi-analytic methods to produce three-dimensional cubes of 21~cm brightness temperature fluctuation at different redshifts. These cubes are represented in rectangular comoving coordinates and can be up to a couple of (Gpc/h)$^3$ in size (e.g. \citet{2011MNRAS.411..955M}), roughly equivalent to $\sim100$~deg$^2$ regions of sky and corresponding depth at the relevant redshifts. In contrast, EoR observations produce three-dimensional data where two of the dimensions map the spatial dimensions of the sky and one dimension measures redshift (frequency), conflating line-of-sight distances with time, also known as the lightcone effect.

In order to match existing 21~cm models to an instrumental observation, we transform the four-dimensional (three spatial and one time) outputs of theoretical 21~cm simulations into a three-dimensional (two angular and one frequency) 21~cm observation cube. Existing techniques for generating lightcone cubes concatenate slices from simulated cubes at multiple redshifts into a single observational cube \citep{2012MNRAS.424.1877D, 2014MNRAS.442.1491D, 2014MNRAS.439.1615Z}. This method captures the time evolution, but not the curvature of the sky. To extend on these methods, we developed a tile-and-grid process that maps a set of simulation cubes from different redshifts to a set of full-sky maps in HEALPix\footnote{\url{http://healpix.jpl.nasa.gov}} coordinates. We elaborate and discuss the procedure in Appendix~\ref{apd:lightcone}. The output from this process is a set of full-sky maps spanning redshifts observed by HERA in steps of 80 kHz spectral channels between $\sim139-195$ MHz, which forms a sky model for this work. We use 80-kHz spectral channel as we adopted the sky model from our previous effort to simulate the MWA. We decided to keep this bandwidth because extensive computing time would be required to rerun the lightcone tiling to match 100-kHz spectral channel, and a 20-kHz increase in channel bandwidth will only improve the thermal noise by $\approx10\%$. Besides, final measurements are usually taken after averaging multiple spectral channels into a larger observed bandwidth to gain additional sensitivity, which we cover in Section~\ref{sec:bandwidth}. 

\subsection{One-point Statistics}
\label{sec:ref_stats}
In this work, we focus on the variance ($S_2$), skewness ($S_3$) and kurtosis\footnote{In statistics, the precise term of this definition is excess kurtosis, which subtracts 3 from the standard definition of kurtosis, $m_4 / (m_2)^2$, to yield zero kurtosis for Gaussian distribution. Here, we simply use kurtosis to refer to excess kurtosis.} ($S_4$), which are standardisations of the 2nd, 3rd and 4th statistical moments ($m_2$, $m_3$ and $m_4$),
\begin{gather}
    S_2 = m_2, \\
    S_3 = m_3 / (m_2)^{3/2}, \\
    S_4 = m_4 / (m_2)^2 - 3,
\end{gather}
where, 
\begin{gather}
    m_p = \frac{1}{N_{pix}}\sum_{i=0}^{N_{pix}} (x_i - \overline{x})^p, \label{eq:moments}
\end{gather} denoted a p-th order statistical moment for a map with data values $x_i$ and mean value $\overline{x}$.

Along with the mean, these three quantities describe deviations in the shape of the brightness temperature PDF relative to a standard Gaussian PDF. The variance measures the spread of the PDF. Both skewness and kurtosis describe the outliers, or the tails, of the PDF in different ways. Skewness measures asymmetry of the outliers, in which positive or negative skewness values indicate that the values of the outliers are greater or less than the mean value of the distribution. Kurtosis describes the tailedness, or density of the outliers; a positive kurtosis indicates more outliers whereas a negative kurtosis indicate fewer outliers. A perfect Gaussian PDF has zero skewness and kurtosis.

\begin{figure}
    \includegraphics[width=0.47\textwidth]{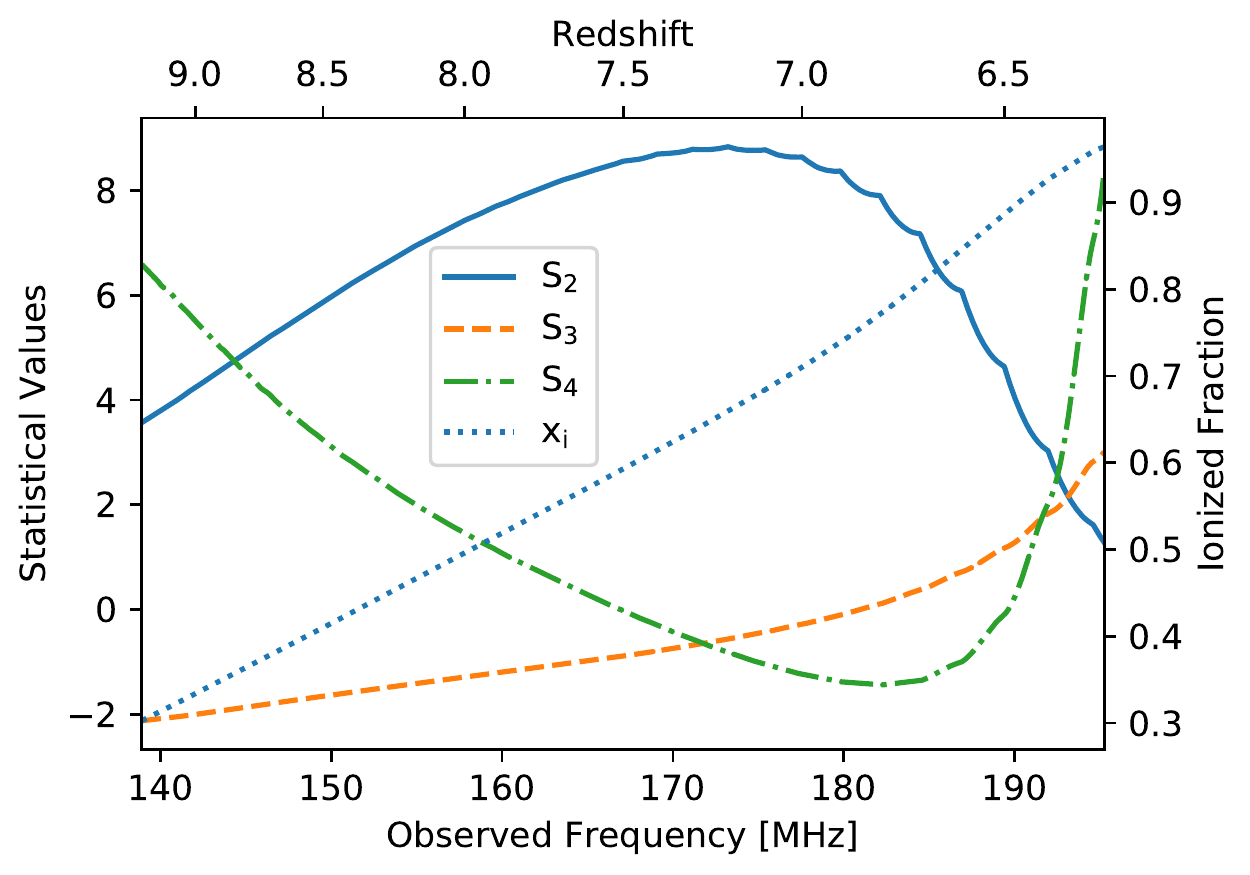}
    \caption{Variance (solid line), skewness (dashed line) and kurtosis (dot-dash line) of the input lightcone model calculated using all of its pixels as a function of frequency and redshift.  The ionised fraction of the model at each redshift is shown as the dotted line. The left y-axis shows corresponding statistical values, whereas the right y-axis shows the ionised fraction. These measurements illustrate the redshift evolution of one-point statistics and act as references for our analysis.}
    \label{fig:model_stats}
\end{figure}
\begin{figure}
    \includegraphics[width=0.47\textwidth]{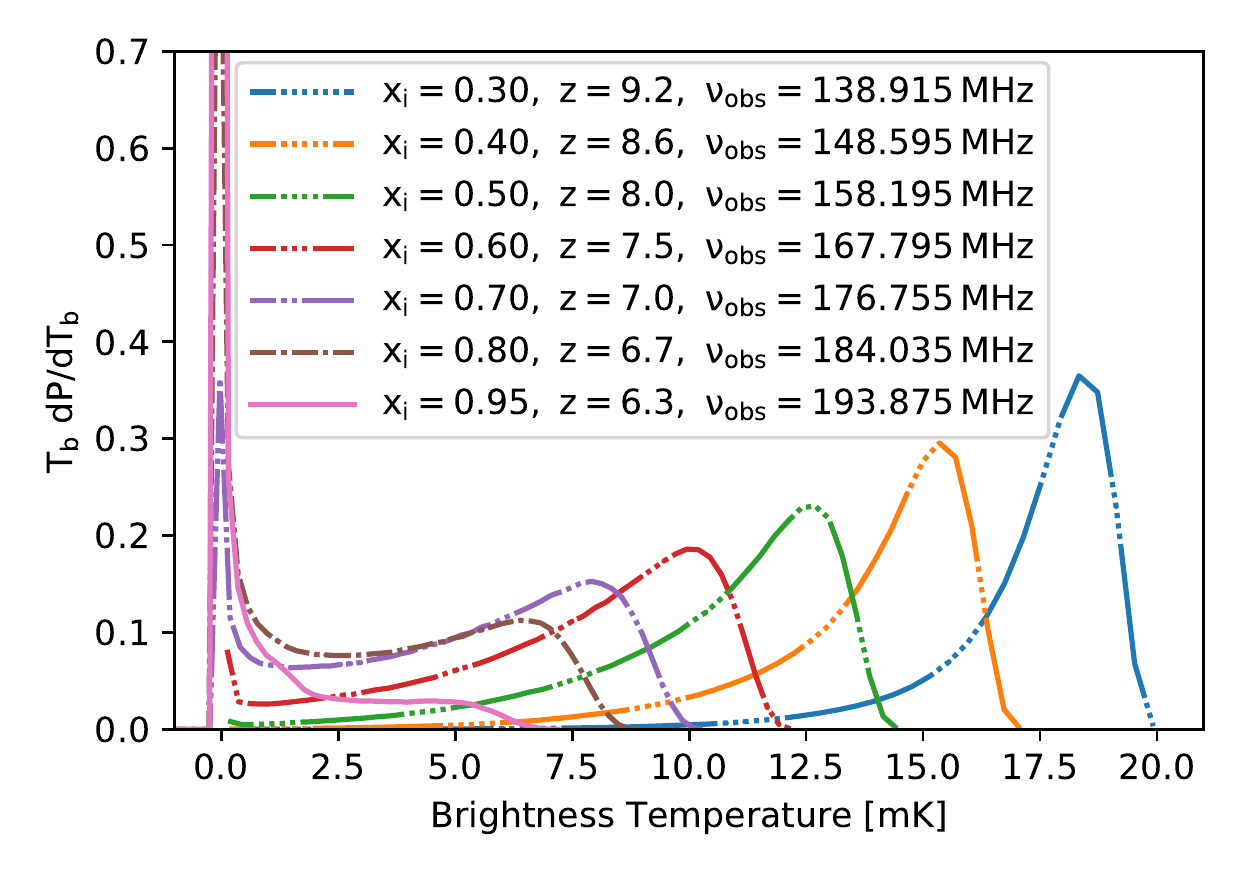}
    \caption{Brightness temperature PDF of the 21~cm lightcone sky model from $x_i=0.3$ to $0.8$, at every $0.1$ step, and at $x_i=0.95$ ionised fraction. The shape of the PDF changes from Gaussian-like to bi-modal to Delta-like function as reionization progresses and manifest the redshift evolution of the statistics shown in Figure~\ref{fig:model_stats}.}
    \label{fig:model_pdf}
\end{figure}

As a reference, we show in Figure~\ref{fig:model_stats} the one-point statistics calculated from our full-sky lightcone, using all of its pixels (without smoothing to HERA's resolution), along with the ionised fraction of the model, as a function of frequency and redshift. In conjunction, we show in Figure~\ref{fig:model_pdf} the PDF of our full-sky lightcone at several ionization fractions to illustrate how the unique evolutions of these statistics arise. Early in reionization ($x_i<0.5$), the brightness temperature fluctuation is relatively Gaussian with small pockets of cold spots that are growing into ionised regions, resulting in a Gaussian-like PDF with a long tail toward zero and hence a negative skewness and a positive kurtosis. As reionization progresses ($x_i\approx0.7$), ionised regions start to dominate the underlying fluctuation, shifting the PDF to a bi-modal. A qualitative interpretation of skewness and kurtosis of a bi-modal distribution is complicated and unintuitive, but this transitional period can be roughly identified by the peak of the variance and the dip of kurtosis between its two zero crossings. As more ionised regions form, the density of the peak of ionized pixels increases. At the very end of reionization, when most of the sky exhibits no 21~cm signal and only a few isolated pockets of emission remain, the PDF becomes a Delta-like function, centring at zero with a long tail toward the warmer temperature, resulting in high skewness and kurtosis. Although we do not explicitly show it, the statistics derived from the lightcone match those of the input 21~cm simulation cubes. The scalloped shape of the variance curve in Figure~\ref{fig:model_stats} is due to the interpolation between theoretical 21cm simulation redshifts. The scalloping is largely removed when the maps are smoothed to the resolution of HERA.

When discussing the detectability of one-point statistics in the following sections, it will be convenient to characterise the evolution of the statistics at these three characteristic times. For our sky model, they correspond to frequency ranges of $\lesssim160$~MHz, $\sim170-190$ MHz and $\sim190-195$ MHz respectively. When HERA angular resolution is taken into account, these features will be weakened, and the corresponding frequency might be shifted, but the rise of skewness and kurtosis near the end of reionization along with the negative kurtosis during the bi-modal transition of the PDF will remain as strong indicators of non-Gaussianity in 21~cm signal as we will show in the following sections. 

It is also important to note that images from HERA observations will have zero mean as a radio interferometer cannot measure the total power of the sky. Thus, the PDF derived from actual observations will be centred around zero. With angular resolution smoothing, the resulting PDF will be similarly blurred, but the basic evolution of the one-point statistics will persist as they are measured relative to the mean of the signal.

\subsection{Mock Observations}
\label{sec:obs}

To simulate mock observations of HERA as described in Section~\ref{sec:hera}, we first smooth our sky model with a Gaussian kernel with a FWHM corresponding to the angular resolution of each of the HERA build-out stages. Then, we pre-allocate 200 non-overlapping fields across the sky, project each field to the instrument observed sine coordinates and measure variance, skewness and kurtosis, using only pixels within a radius equal to half of a FWHM of the HERA primary beam from the field centre. Before calculating statistics, each field is subtracted by its mean value to emulate the absence of the mean value of the sky in interferometric observations. Then, we randomly select 20 measurements and use their mean as an estimate for statistics recoverable by a HERA drift scan. To estimate the sample variance of the drift scan, we repeat the random draw of the 20 measurements, calculate the drift scan estimate from each draw, and use the standard deviation of all estimates as the drift scan sample variance uncertainty. In addition, we use the standard deviation of all 200 single-field  measurements as an estimate for the sample variance uncertainty in single field observations. We add the sample variance uncertainty to the thermal noise uncertainty in quadrature to estimate the total uncertainty in any single-field measurement and propagate to drift scan measurements accordingly.  We use statistics calculated from all pixels of the sky model smoothed to HERA angular resolution as the estimate of the ideal signal.

We adopt thermal noise uncertainty calculations from \citet{Watkinson:2014jv}, expanding their derivation to kurtosis.  In summary, the uncertainty in interferometric observations ($\Delta T_n$) can be described by the system temperature ($T_{sys}$) of the array, array filling factor ($\eta_f$), spectral channel bandwidth ($\Delta\nu$) and integration time of the observations ($t_{int}$) \citep{2006PhR...433..181F},
\begin{gather}\label{eq:noise1}
    \Delta T_n = \frac{T_{sys}}{\eta_f\sqrt{\Delta\nu\,t_{int}}}.
\end{gather}
By assuming that the system temperature is dominated by the Galactic synchrotron radiation at the EoR observing frequency, $T_{sys} \approx T_{sky}=180~(\nu/180~\mathrm{MHz})^{-2.6}\mathrm{K}$ \citep{2016MNRAS.455.3890M}, Equation~\ref{eq:noise1} can be expanded to obtain the thermal noise uncertainty ($\sigma_n$) in redshifted 21~cm observations,
\begin{gather}\label{eq:noise2}
\begin{split}
    \sigma_n = 2.9\,\mathrm{mK} 
    &\left(\frac{10^5\,\mathrm{m}^2}{A_{tot}}\right)
    \left(\frac{10\,\mathrm{arcmin}}{\Delta\theta}\right)^2 \\
    &\times \left(\frac{1+z}{10.0}\right)^{4.6}
    \sqrt{\left( \frac{1\,\mathrm{MHz}}{\Delta\nu} \frac{100\mathrm{h}}{t_{int}}\right)},
\end{split}
\end{gather} 
which depends on the total effective collecting area of the array ($A_{tot}$), the angular resolution of the array ($\Delta\theta$), redshift ($z$) and integration time of the observations ($t_{int}$) in addition to the spectral channel bandwidth. 

Thermal noise induced estimator variance of the 2nd ($V_{\hat{m_2}}$), 3rd ($V_{\hat{m_3}}$) and 4th ($V_{\hat{m_4}}$) order statistical moments can be derived from this noise description based on a statistical framework to obtain,
\begin{gather}
    \label{eq:m2_var}
    V_{\hat{m}_2} = \frac{2}{N}
        (2 m_2 \sigma_n^2 + \sigma_n^4), \\
    \label{eq:m3_var}
    V_{\hat{m}_3} = \frac{3}{N}
        (3 m_4 \sigma_n^2 + 12 m_2 \sigma_n^4 
        + 5 \sigma_n^6), \\
    \label{eq:m4_var}
    V_{\hat{m}_4} = \frac{8}{N}
    (2 m_6 \sigma_n^2 + 21 m_4 \sigma_n^4
        + 48 m_2 \sigma_n^6 + 12 \sigma_n^8),
\end{gather}
and later propagated to derive the estimator variance of the signal variance ($V_{\hat{S_2}}$), skewness ($V_{\hat{S_3}}$) and kurtosis ($V_{\hat{S_4}}$), yielding,
\begin{gather}
    \label{eq:var_var}
    V_{\hat{S_2}} = V_{\hat{m}_2} \\
    \label{eq:skew_var}
    V_{\hat{S_3}} \approx \frac{1}{(m_2)^3}V_{\hat{m}_3}
        + \frac{9}{4}\frac{(m_3)^2}{(m_2)^5}V_{\hat{m}_2}
        - 3\frac{m_3}{(m_2)^4}C_{\hat{m}_2\hat{m}_3}, \\
    \label{eq:kurt_var}
    V_{\hat{S_4}} \approx \frac{1}{(m_2)^4} V_{\hat{m}_4}
        + 4 \frac{(m_4)^2}{(m_2)^6} V_{\hat{m}_2} - 4 \frac{m_4}{(m_2)^5} C_{\hat{m}_2\hat{m}_4},
    \end{gather}
where $N$ is the number of samples in a measurement. $C_{\hat{m}_2\hat{m}_3}$ and $C_{\hat{m}_2\hat{m}_4}$ are the estimator covariance of the relevant moments, which can be derived in the same manner to obtain,
\begin{gather}
    C_{\hat{m}_2\hat{m}_3} = \frac{6}{N}m_3\sigma_n^2, \\
    C_{\hat{m}_2\hat{m}_4} = \frac{4}{N}
        (2 m_4 \sigma_n^2 + 9 m_2 \sigma_n^4
        + 3 \sigma_n^6).
\end{gather}
The thermal noise uncertainty of one-point statistics of each single field is then just a square root of its estimator variance. Appendix~\ref{apd:noise} provides a more detailed summary of \citet{Watkinson:2014jv} procedure as well as our derivation of the estimator variance of the 4th statistical moments and the kurtosis.

We assume $t_{int}=100$~hours for every $\Delta\nu=80$~kHz spectral channel as discussed earlier. The number of samples per measurement in an observation will be limited by the field of view and the angular resolution of the telescope. Because our maps oversample the angular resolution with multiple pixels, we calculate the number of independent resolution elements per pixel from the ratio of the pixel and the resolution element areas, $f_{s}=\Delta\Omega_{pix} / \Delta\Omega_{res}$, and multiply this factor to the number of pixels in a measurement to obtain $N$. The oversampling of the angular resolution does not affect the statistics because the over sampling factor is cancelled out in the moment equation (see Equation~\ref{eq:moments}).

Table~\ref{tab:sim} summaries the parameters of our mock observations, where we refer to HERA240 and HERA350 arrays without their outriggers as HERA240 Core and HERA350 Core respectively.

\begin{table*}
\begin{center}
\begin{tabular}{l c c c c c}
\hline
Simulation Parameters &HERA19  &HERA37 &HERA128 &HERA240 Core &HERA350 Core \\
\hline
Collecting area (m$^2$)   	&2,925		&5,696     &19,550		&33405		&50,953     \\
Maximum baseline (m)     &70		&98     	&182 	&238	 &294       \\
Angular Resolution (deg) &$\sim1.53-2.16$ 		&$\sim1.10-1.54$ 	&$\sim0.59-0.83$  &$\sim0.45-0.63$ 	&$\sim0.37-0.51$ \\
\hline
\end{tabular}
\caption{\label{tab:sim}Instrument specifications for the HERA build-out stages used in our mock observations. We perform the simulation over$\sim$56 MHz bandwidth, from $\sim$139-195~MHz, with 80~kHz spectral channel bandwidth. Further information on the array configurations can be found in \citet{2017PASP..129d5001D}.}
\end{center}
\end{table*}

\section{Results}
\label{sec:results}
In this section, we present one-point statistical measurements from the simulated observations. We will first show images of the mock observations and present measurements from the HERA350 Core simulation with 80 kHz channel bandwidth. Then, we will investigate if sensitivity could be further improved with bandwidth averaging.

\subsection{Simulated Observations}
\begin{figure*}
    \includegraphics[width=1.0\textwidth]{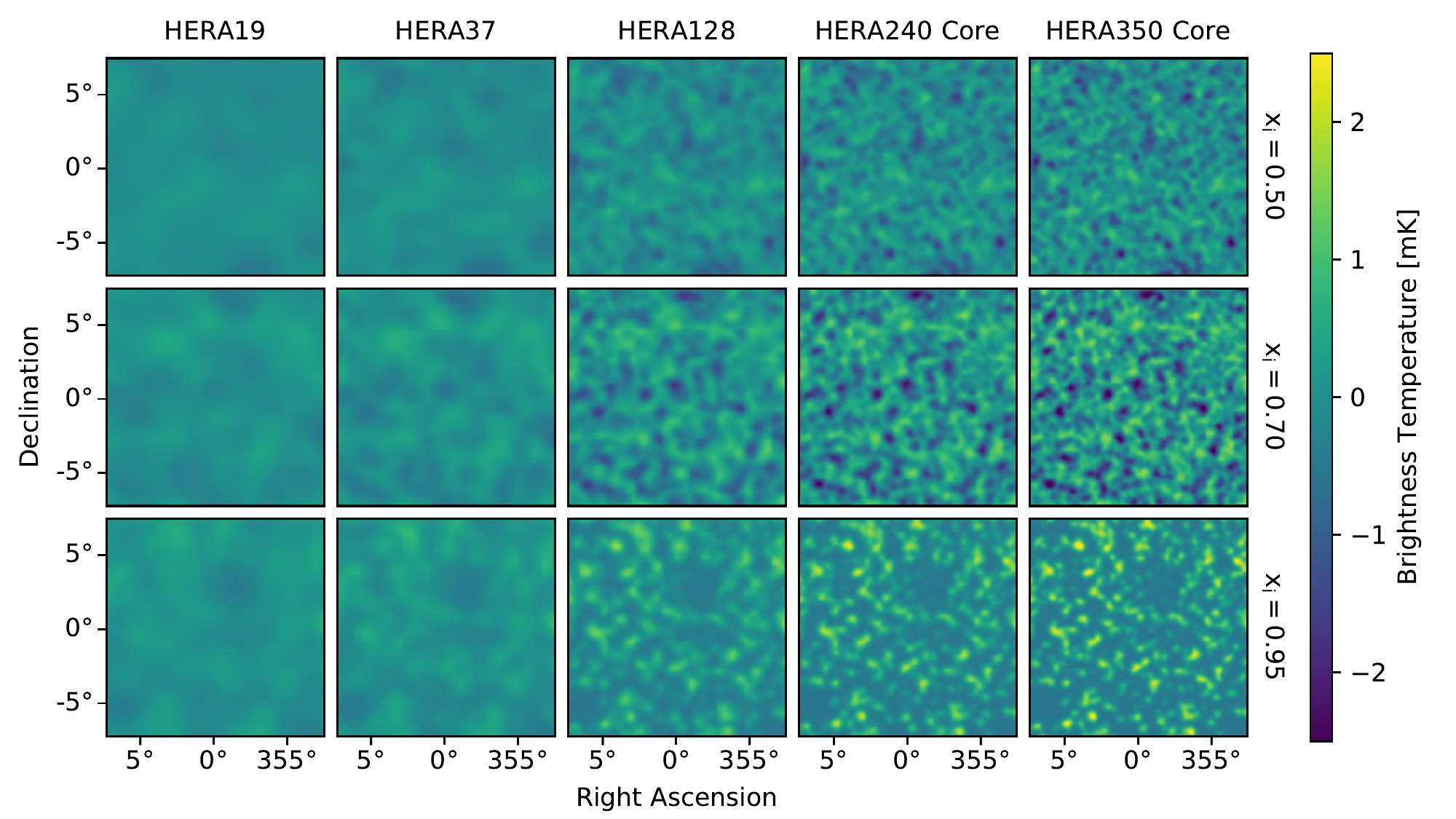}
    \caption{Simulated observations at ionised fraction of 0.5, 0.7 and 0.95 (top to bottom rows) with different HERA build-out stages (left to right columns). The 21~cm lightcone model is smoothed to the resolution of each array, showing more pronounced fluctuations as the angular resolution increases. No thermal noise is included in the simulation.}
    \label{fig:maps}
\end{figure*}

\begin{figure*}
    \includegraphics[width=1.0\textwidth]{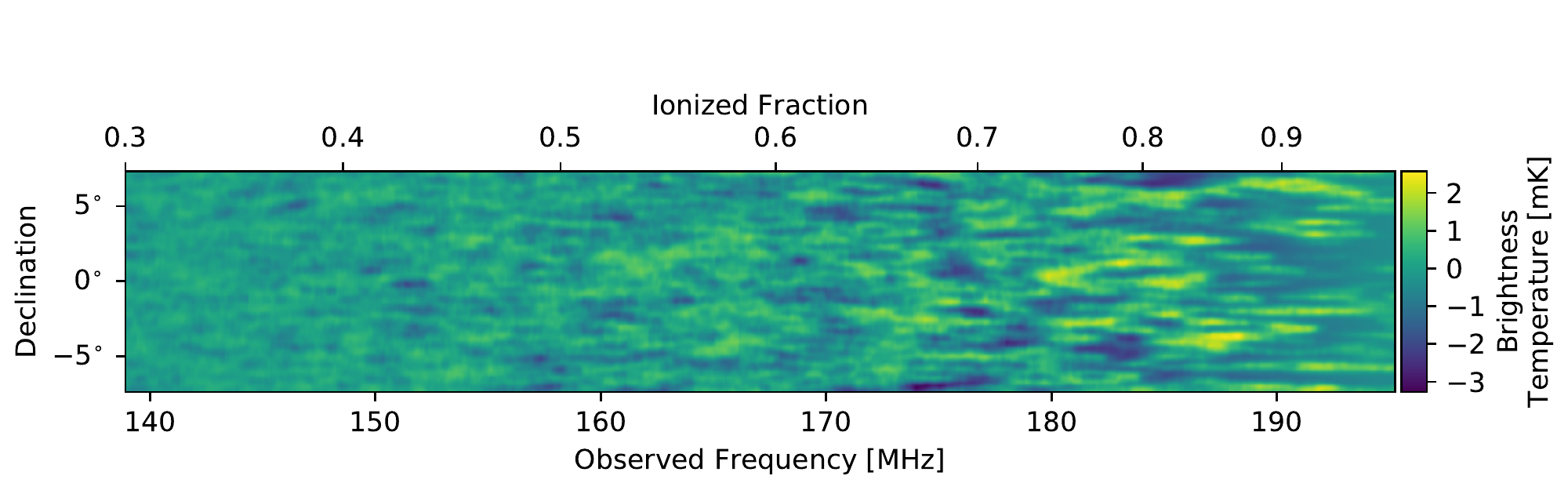}
    \caption{Lightcone slice from the HERA350 Core mock observations. The size scale along the frequency direction grows as reionization progresses and reaches a typical size of $\sim4$ MHz near the end of reionization.}
    \label{fig:lightcone}
\end{figure*}

Figure~\ref{fig:maps} shows simulated observations at $x_i=0.5$, 0.7 and 0.95 for all of the HERA build-out stages with all panels taken from the same field in our sky model. The angular resolution increases as the telescope grows and the fluctuations become more pronounced. Figure~\ref{fig:lightcone} shows a cut along the frequency direction from the same field at HERA350 Core angular resolution to illustrate the lightcone evolution. The size scale along the frequency direction grows as reionization progresses and reaches a typical size of $\sim4$ MHz near the end of reionization. The brightness temperature scale of the  observed maps is centred around zero with both negative and positive values due to the lack of mean measurements. 


\subsection{HERA350 Core Measurements}
\label{sec:hera350_stats}
\begin{figure}
    \includegraphics[width=0.48\textwidth]{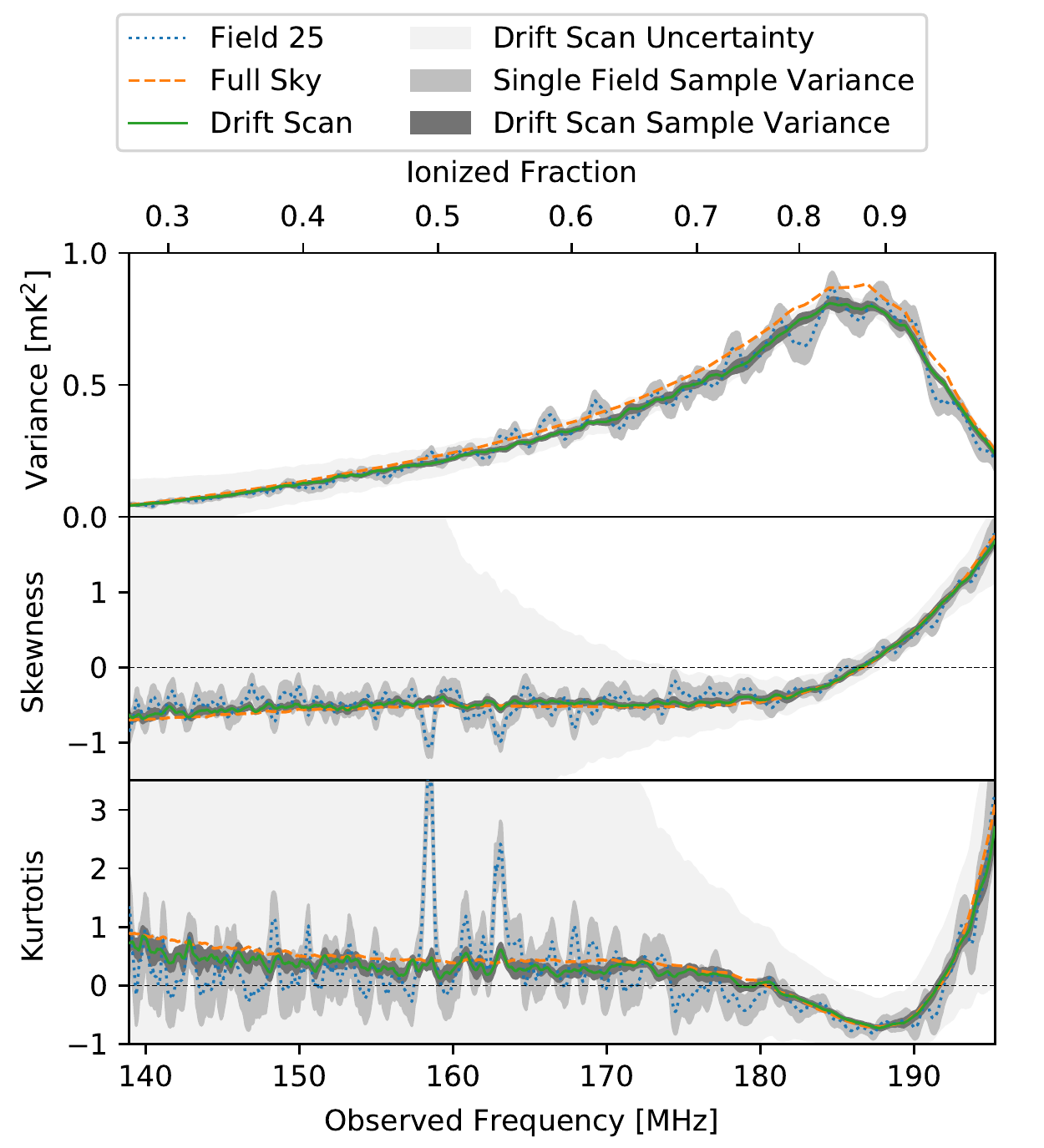}
    \caption{Variance, skewness and kurtosis (top to bottom panels) measured from simulated HERA 350 Core observations with 80 kHz channel bandwidth as a function of observing frequency and ionised fraction. The dotted line shows an example of statistics measured from a single field. The solid line shows the drift scan measurements, defined as the mean of 20 single-field measurements. The dashed line shows statistics derived from the full sky after smoothing to HERA350 Core angular resolution as a reference. The three shaded regions shows sample variance uncertainty of the drift scan measurements (dark), which is only visible in the kurtosis, sample variance uncertainty of a single field measurement (medium), and the uncertain of the drift scan measurements propagated from the single field sample variance and thermal noise uncertainty (light). HERA350 Core will be sensitive to one-point statistics, particularly in the second half of reionization. Skewness and kurtosis measurements are limited by thermal noise, but detections are possible near the end of reionization, while both thermal noise and sample variance are insignificant in variance measurements.}
    \label{fig:stats_hera350}
\end{figure}

Figure~\ref{fig:stats_hera350} shows one of the drift scan measurements derived from the HERA350 Core simulation with 80 kHz bandwidth, an example of single field measurements, and statistics of the sky model smoothed to HERA350 Core angular resolution. Comparing the drift scan measurements with the input model statistics in Figure~\ref{fig:model_stats}, we see the effect of HERA's relatively poor angular resolution, which acts to smooth the input maps, reducing the observed variance, skewness and kurtosis. We will see below that this effect is especially pronounced for the HERA build-out phases, where the coarser angular resolution is predicted to yield one-point statistics that are only weakly non-Gaussian \citep{2015MNRAS.449L..41M}. 

Nevertheless, it is evident that HERA350 Core will be sensitive to one-point statistics, particularly in the second half of reionization where the rise of skewness and the pronounced negative dip and rise of kurtosis indicate non-Gaussian fluctuations. Uncertainty from thermal noise in the drift scan measurement of the variance is small throughout the redshift ranges, becoming negligible in comparison to the contribution from sample variance at the end of reionization, and should allow high sensitivity measurements. For skewness and kurtosis, the thermal noise is significant, resulting in large uncertainty in the drift scan measurements at the beginning of reionization, but the thermal uncertainty decreases as frequency increases, becoming minimal at the end of reionization, beyond $\sim$170 MHz for our sky model. The result is that the variance measurement is limited by thermal noise early in reionization and later by sample variance near the end of reionization while skewness and kurtosis measurements will be dominated by thermal noise throughout reionization, but significant detections will be possible near the end of reionization where the signal is strong.

As expected, statistics from single field measurements exhibit fluctuations due to sample variance as opposed to the smoothly evolving ensemble statistics derived from the full-sky model. Occasionally, the fluctuations sharply rise despite the overall small sample variance uncertainty. These strong fluctuations are interesting in their own right, and we explore their behaviour in Section~\ref{sec:kurtosis}. The 20-field averaged, drift scan observations provide a much more faithful recovery of the model statistics with negligible fluctuations from sample variance.

\subsection{Improving Sensitivity with Bandwidth Averaging}
\label{sec:bandwidth}
We have only considered measurements from mock observations with 80 kHz channel bandwidth in Section~\ref{sec:hera350_stats}. The narrow channel bandwidth limits thermal noise performance in the measurements.  In this section, we introduce two methods of bandwidth averaging, a commonly used ``frequency binning'' and a less-common method, mainly used in the 21 cm power spectrum measurements that we term ``frequency windowing.''

Frequency binning improves thermal noise uncertainty and can be done by averaging maps of neighbouring spectral channels to produce a single output map with larger effective channel bandwidth and lower thermal noise.   This is an effective strategy to improve thermal noise until the bin size becomes larger than the typical size of the features in the signal, at which point further increasing the bin size will diminish the strength of the signal by averaging over uncorrelated regions. 

We find in Section~\ref{sec:hera350_stats} that skewness and kurtosis measurements are limited by thermal noise. We will show below that frequency binning can be used to significantly improve the sensitivity of HERA to these two statistics.  We also find in Section~\ref{sec:hera350_stats} that measurements of the variance by HERA will be constrained by sample variance at high frequencies toward the end of reionization, rather than by thermal noise. Frequency binning will not improve sample variance sensitivity because the number of of samples that can be used in a measurement is the same for both the binned and un-binned maps when observing over the same field with the same instrument. Thus, it is of interest to explore an additional approach aimed at reducing sample variance. Since sample variance depends on the number of samples per map, which is fixed by the angular resolution and the field of view, the only way to increase the number of samples used in an estimate is to use maps from multiple spectral channels as a single data set. In this method, we form a three-dimensional data cube from multiple maps of neighbouring spectral channels and measure the statistics of the cube, using all samples within the field of view. We term this process frequency windowing. Thermal noise per pixel is unchanged with frequency windowing because the native spectral channel is preserved, but thermal noise uncertainty on the one-point statistic estimates will still decrease due to the  $1/N$ factor in the estimator variance equations (see Equation~\ref{eq:m2_var}--\ref{eq:m4_var}).  

To explore the relative trade-offs between the two methods, we perform frequency binning and frequency windowing on the brightness temperature maps from our HERA350 Core simulation, varying bin and window sizes from 1-8~MHz with 1~MHz increment. We start from the highest spectral channel in our $\sim$60-MHz bandwidth, at 195 MHz, and bin or window, progressing down to lower frequencies. The process is repeated on all 200 sample fields. Then, we make drift scan measurements and estimate the uncertainties for each case using the same method as described in Section~\ref{sec:sim}.

The observed variance decreases across the observed frequency range when frequency binning is used, with particularly rapid decline near the variance maxima. This is expected as the signal strength along the frequency dimension in our HERA350 Core mock observations, shown in Figure~\ref{fig:lightcone}, is smaller and more uniform early in reionization but grows and reaches the maximum strength and maximum variance at around 185 MHz. Binning the signal when both the signal strength and signal variance are high will greatly reduce the signal amplitude, resulting in much smaller observed variance. As a side note, the observed variance only slightly declines between the native 80-kHz channel bandwidth and 1-MHz binning case. This is expected as the coherence length of the 21~cm signal is predicted to be approximately 1~MHz \citep{2005ApJ...625..575S}. Thus, the bin size finer than 1 MHz would not dramatically alter the signal variance. Although not shown, we find that the overall evolution of skewness and kurtosis recovered by the instrument do not change when frequency binning is used apart from slight variations between bins.

Compared to frequency binning, frequency windowing simply adds more samples along the frequency dimension. With no averaging of maps from neighbouring spectral channels, variance of the signal within each spectral channel is preserved, and the observed variance would be the mean value of variance of all spectral channels within that window. Skewness and kurtosis are also preserved as the added samples from neighbouring spectral channels do not alter the shape of the PDF but contribute to form a more well sampled PDF, unless strong redshift evolution occurs between spectral channels. All statistics measured from frequency windowing data also show less variation between window to window and follow the sky model more closely.

\begin{figure*}
    \includegraphics[width=1.0\textwidth]{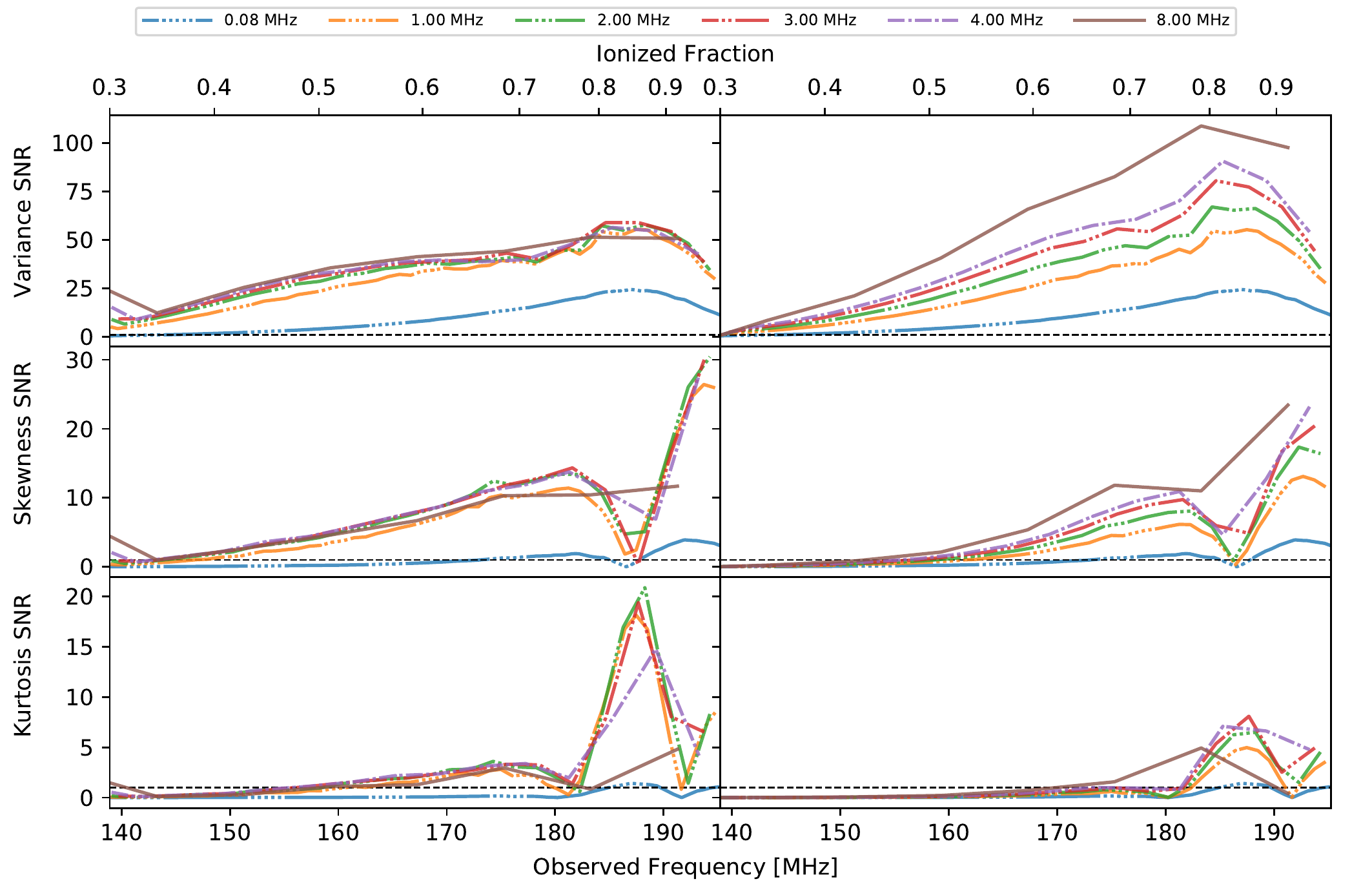}
    \caption{Comparing SNR measured from HERA350 Core simulation with different frequency binning (left column) and frequency windowing (right column) cases. SNR is defined as the ratio of the absolute values of the mean statistics and uncertainty from drift scan observations. Frequency windowing improves the sensitivity more than frequency binning at nearly all redshifts in the variance measurements due to the reduced sample variance, whereas frequency binning is more effective for skewness and kurtosis measurements as HERA observations on these two statistics are limited by thermal noise. The horizontal dashed lines indicate SNR=1}
    \label{fig:snr_bw_hera350}
\end{figure*}

We find that frequency binning is more effective for the skewness and kurtosis measurements at nearly all redshifts in our simulations since sample variance dominates the uncertainty over much of the modelled band, whereas frequency windowing is better for the variance measurements. To illustrate, we compare in Figure~\ref{fig:snr_bw_hera350} the signal-to-noise ratio (SNR) of the statistics for different binning and windowing cases measured from the HERA350 Core simulation. We define the SNR as the absolute values of statistics divided by their uncertainties, and we use the mean values of the measurements from 10 independent sets of the drift scan simulations in the calculation to more clearly illustrate the results. The SNR improvement from frequency binning in the variance measurements saturates after the bin size reaches a few MHz because thermal noise has been reduced below sample variance at that point, putting the measurements in sample variance limited regime. The binned signal amplitude also declines at approximately the same rate as the thermal noise uncertainty when the bin size is larger than the typical signal feature size. With frequency windowing, SNR of the variance measurements continues to improve with larger windows. Although we only investigate up to an 8-MHz window case, we expect the SNR of the variance to continue improving beyond the 8 MHz window size. However, redshift evolution will have detrimental effect on the signal with that large bandwidth; thus, the logical choice would be to perform frequency windowing on the variance measurements only with a large enough window size to obtain sufficient SNR.

A general conclusion is that frequency windowing should be used in observations that are sample variance dominated, whereas frequency binning maybe more beneficial when observations are thermal noise dominated (depending on the spectral coherence of the underlying signal). This statement is easiest to explain quantitatively. For an observation at a particular frequency, integration time and angular resolution, the thermal noise description from Equation~\ref{eq:noise2} can be plugged into the estimator variance formulas in Equation~\ref{eq:m2_var}, \ref{eq:m3_var} and \ref{eq:m4_var} to obtain a simplified thermal noise uncertainty equation for $p$-th order one-point statistics, 
\begin{gather}
    \sigma_{\hat{m}_p} = \sqrt{V_{\hat{m}_p}} \propto \frac{\sigma_n^{p}}{\sqrt{N}} \propto \frac{1}{\sqrt{N(\Delta \nu)^p}}.
\end{gather}
When performing frequency binning with $n_{ch}$ channels, the number of samples in a measurement $N$ is unchanged, but the frequency binned map will have an effective channel bandwidth of $\Delta \propto n_{ch}$, resulting in $\sqrt{n_{ch}^p}$ reduction of thermal noise uncertainty. In contrast, performing frequency windowing with $n_{ch}$ channels will not alter the channel bandwidth, but the number of samples in a measurement is multiplied by $n_{ch}$, reducing thermal noise uncertainty by just $\sqrt{n_{ch}}$. However, the increased number of samples from frequency windowing will also reduce sample variance, allowing detection with higher sensitivity in the case where thermal noise is no longer the limiting constraint of the observations.

For HERA, frequency binning will yield sufficient sensitivity on all observations of one-point statistics. Frequency windowing could also be used in the variance measurement if higher sensitivity is desired. A combination of both method should also be possible to further improve the sensitivity but is beyond of scope of the study in this work.

\subsection{Performance of HERA Build-out Stages}
\label{sec:performance}

Although it is clear from Section~\ref{sec:hera350_stats} and \ref{sec:bandwidth} that the complete HERA 350 Core array will be able to mitigate sample variance by averaging statistics measured over multiple fields across the sky and utilising bandwidth averaging, it is important to see how the smaller, planned built-out, arrays will perform as data from HERA 350 Core will not be available until after 2020. 

In general, smaller arrays have two key disadvantages of less collecting area and worse angular resolution. Typically, smaller collecting area increases the thermal uncertainty, while poorer angular resolution smooths over the intrinsic 21~cm features, yielding a more-Gaussian like signal with lower amplitude than for higher angular resolutions. However, HERA is very close to being a filled aperture array, resulting in the angular resolution and array collecting area that roughly scale as the maximum baseline and the maximum baseline squared, respectively. As a consequence, it can be shown from Equation~\ref{eq:noise2} that the thermal uncertainty per resolution element of HERA does not change with the size of the array. Thus, the only disadvantage of smaller HERA array is the reduced angular resolution that damps the signal and lowers the number of independent resolution elements per map.

For a detailed investigation, we perform frequency binning and frequency windowing on the mock observations of all HERA build-out stages, and calculate the drift scan statistics, sample variance uncertainty and thermal noise uncertainty for each case following the methods  described in Section~\ref{sec:obs}. The HERA build-out arrays cover angular resolution from $\sim$1.5$^{\circ}$ to 0.5$^{\circ}$ as the array grows. Thus, this study also gives an insight into the effects of angular resolution on the one-point statistics.

\begin{figure*}
    \includegraphics[width=1.0\textwidth]{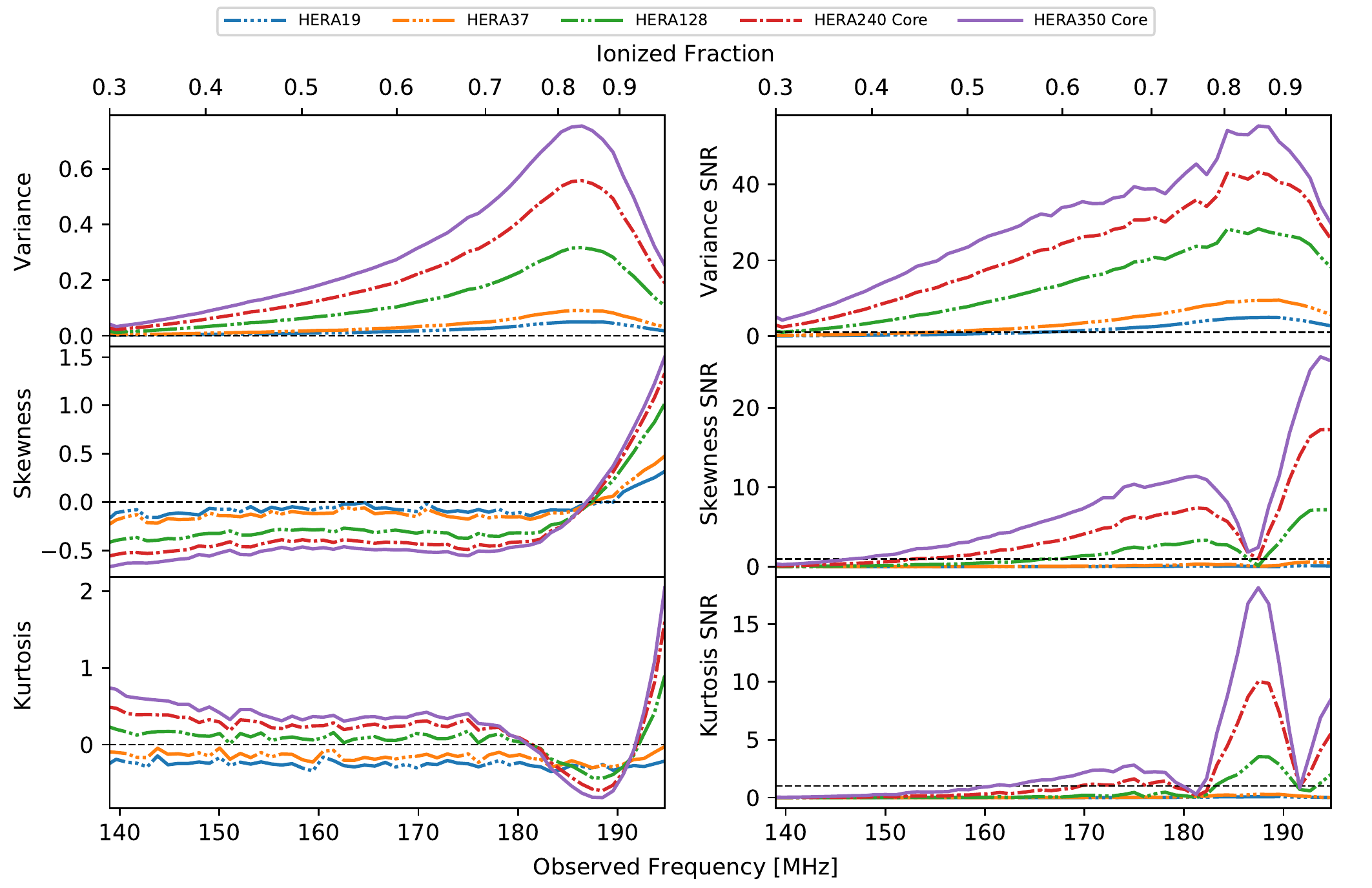}
    \caption{Mean drift scan statistics (left) and SNR (right) measured from the mock observations of all planned built-out HERA stages with 1-MHz frequency binning applied before the calculations. All HERA configurations should be able to measure the variance with high SNR. In addition, HERA128 and above will be able to measure the characteristic rise of skewness and the dip, then rise, of kurtosis near the end of reionization with sufficient SNR. The horizontal dashed lines on the statistics and SNR columns indicate zero statistical values and SNR=1 respectively.}
    \label{fig:4MHz_snr_heraxx}
\end{figure*}

Figure~\ref{fig:4MHz_snr_heraxx} shows the one-point statistics measured from the mock observations of each of the build-out stages, along with the corresponding SNR calculated in the same manners as in Section~\ref{sec:bandwidth}, with 1 MHz frequency binning applied before the calculations. It is clear that the derived statistics are affected by the increasing angular resolution as the array grows.  Variance decreases with smaller arrays, similar to the effect of frequency binning. Skewness and kurtosis measured from smaller, HERA19 and HERA37, arrays also vanish, only fluctuating near zero throughout much of reionization.  In contrast, the larger HERA128, HERA240 Core and HERA350 Core arrays exhibit non-zero skewness and kurtosis even early in reionization, diverging more from zero as the angular resolution improves. The negative region of kurtosis near the end of reionization only reaches significance in observations with these large build-out phases.

This resolution effect is an expected consequence from the angular resolution smoothing. With finer angular resolution, larger HERA arrays can resolve more of the intrinsic underlying fluctuation and preserve the amplitude of the signals. The shape of the PDF distribution of the signal is also preserved, and thus the values of skewness and kurtosis remain closer to the intrinsic level of the signals. When the angular resolution becomes larger than the typical sizes of the underlying signals, angular resolution smoothing blurs out the fluctuations, reduces the overall signal amplitude and shift the PDF from non-Gaussian to Gaussian. As a result, variance is greatly reduced, and skewness and kurtosis vanish. These results are in agreement with \citet{2015MNRAS.449L..41M}, who suggests, based on the study of 21~cm power spectrum SNR, that 21~cm signal would only be weakly non-Gaussian at a degree scale near the end of reionization, comparable to angular resolution of HERA19 and HERA37 arrays, and be mostly Gaussian otherwise.

Apart from better resolving the intrinsic fluctuations, increasing the angular resolution will also reduce sample variance. All statistics measured from the mock observations of the arrays with finer angular resolutions also show less variation between bin to bin and follow the sky model more closely.

Our simulations suggest that all HERA configurations should be able to measure the variance with high SNR. In addition, HERA128, HERA240 Core and HERA350 Core will be able to measure the characteristic rise of skewness and the dip, then rise, of kurtosis near the end of reionization with sufficient SNR, especially when frequency binning is used, as demonstrated in Figure~\ref{fig:4MHz_snr_heraxx}. 

\section{Individual Field Observations}
\label{sec:kurtosis}
\begin{figure}
    \includegraphics[width=0.47\textwidth]{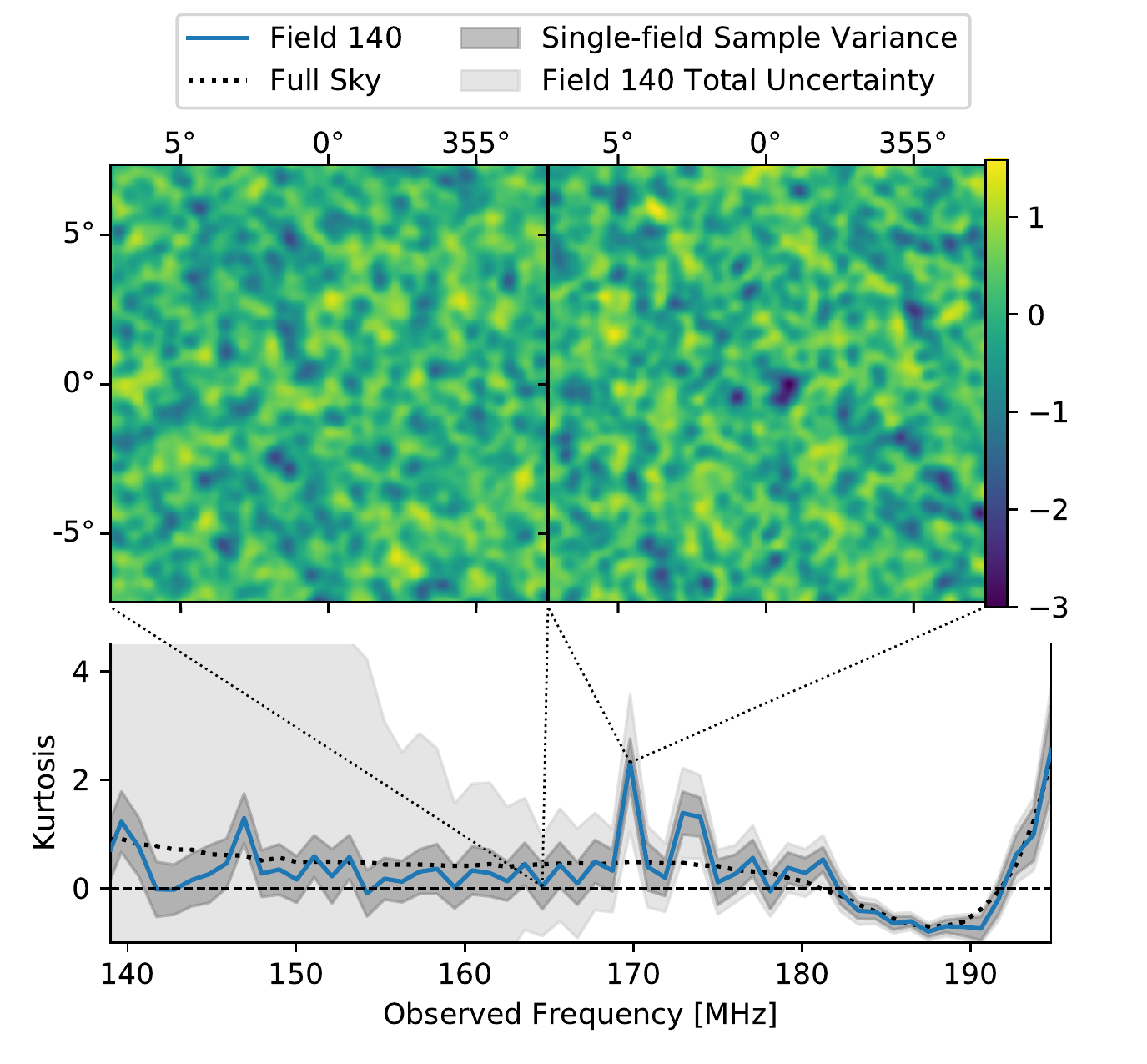}
    \caption{Illustrating the correlation between strong sample variance fluctuation in the single-field measurement of kurtosis and the outlying cold or hot spots in the observed field. The bottom panel shows the kurtosis measured from field 140 from HERA350 Core simulation with 1-MHz frequency binning applied (solid line), overlaid on top of a 1-$\sigma$ single-field sample variance uncertainty (dark shade) and total uncertainty (light shade), and the expected full sky kurtosis (dotted line). The top two panels show maps of the underlying brightness temperature signal at $\sim$170 MHz, where the kurtosis rises above the sample variance uncertainty (right), and at $\sim$165 MHz, where the kurtosis is near zero (left). The outlying cold spots perturbs the tail of the PDF, causing kurtosis to rise. This statistical feature could potentially be used as a bubble indicator.}
    \label{fig:kurt_maps}
\end{figure}

In this paper, we focus on recovering the one-point statistics over multiple fields corresponding the full drift scan area that will surveyed by HERA.  Here we briefly turn our attention to the individual 9x9 deg$^2$ HERA beam fields in our simulations.  As expected, statistics measured within individual fields are more susceptible to sample variance.  For example, there is a strong kurtosis spike near 170 MHz in the measurement from field number 140 from our HERA350 Core simulation when 1-MHz frequency binning applied.  This kurtosis spike is a factor of $\sim 5$ above the sample variance expected for the field.   Significant outlier deviations such as these are seen in fields with one or two large cold or hot spots dominating the underlying fluctuation. Figure~\ref{fig:kurt_maps} illustrates the case.  When cold or hot spots appear in the observed fields, they perturb the PDF, adding more density to the tails of the distribution, and causing kurtosis to rise. In some occasions, these outliers also shift the symmetry of the PDF and cause strong troughs in the skewness as appeared in Figure~\ref{fig:stats_hera350}. Outlier one-point statistics of small fields may provide a simple and robust bubble detector and a new tool to constrain reionization models.  Statistical study of the frequency of occurrence of the outlier kurtosis (or skewness) peaks could be conducted for a given instrument and related to predictions from reionization models.  The outliers should be less susceptible to noise in the underlying fluctuations than typical estimates of the statistics.  Investigation of this possible metric would require further study that is beyond the scope of this work.

\section{Conclusion}\label{sec:conclusion}
We have established a baseline sensitivity analysis of 21~cm one-point statistics from the EoR for HERA through realistic mock observations of redshifted 21~cm brightness temperature fluctuations. 

We develop a tile-and-grid method that transforms a suite of small 21~cm simulation cubes into a full-sky lightcone input model that matches the dimensionality of the observational data. We incorporate the angular resolution effects of all of the planned build-out stages of HERA to gauge the sensitivity of the array as it grows in size. We use simple Gaussian smoothing to incorporate the array angular resolution, using multiple kernel sizes that match angular resolutions of the different build-out stages. The span of resolutions also allows us to study their effects on the statistics. Apart from the variance and skewness that have extensively been covered in previous studies, we also measure kurtosis from our mock observations as well as deriving sample variance uncertainty associated with the measurements. Uncertainty from thermal noise is mathematically derived from the framework developed in \citet{Watkinson:2014jv}, where we have extended their derivation to kurtosis. We calculate SNR to gauge the sensitivity of the measurement and perform frequency binning and frequency windowing to investigate if the sensitivity can be further improved. We ignore foreground contamination and other systematics in this work, postponing them to future works.

Our results show that measurements of 21~cm one-point statistics by HERA will be sample variance limited throughout reionization for the variance measurements while skewness and kurtosis sensitivities will be limited by thermal noise. Frequency binning can be used in all measurements to improve the sensitivity. In addition, frequency windowing can be used in the measurements, particularly in the variance, to further improve the sensitivity once thermal noise has been reduced below the uncertainty from sample variance. However, care must be taken to not use a bin or a window size that is too large to avoid redshift evolution. In addition, all build-out stages of HERA will be able to measure variance with high sensitivity, and HERA128 and above will also be able to measure skewness and kurtosis.

An introduction of kurtosis into our analysis has led us to identify kurtosis peaks as potential indicators of outlying cold or hot spots in individual fields of observations. Kurtosis will sharply rise when a few hot or cold outlying regions appear on top of the underlying Gaussian-like signal in the observed field. Further investigation of this feature is beyond the scope of this work.

Although we only focus on HERA in this paper, our results should be applicable to future arrays such as the SKA. In particular, the thermal uncertainty will be lower and the signal strength and sample variance uncertainty will be further improved due to its higher angular resolution.

There are three main obstacles in pursuing future detection and characterisation of the EoR with 21~cm one-point statistics: (1) the sensitivity of existing reionization arrays to the statistics, (2) mitigation of astrophysical foreground contamination, and (3) estimation of reionization parameter constraint from the statistics. We have extensively covered the first point in this paper, focusing on the HERA array, and we are actively investigating the effects of foreground contamination on our results. The last point is arguably the least well-understood, and we will investigation the topic in the future.

\section*{Acknowledgements}
This work is supported by the U.S. National Science Foundation (NSF) award AST-1109156. P.K. is supported by the Royal Thai Government Scholarship under the Development for Promotion of Science and Technology Talent Project. D.C.J. is supported by an NSF Astronomy and Astrophysics Postdoctoral Fellowship under award AST-1401708. A.P.B. is supported by an NSF Astronomy and
Astrophysics Postdoctoral Fellowship under award AST-1701440. N.T. is a Jansky Fellow of the National Radio Astronomy Observatory.




\bibliographystyle{mnras}
\bibliography{references}




\appendix

\section{Full-sky Lightcone}
\label{apd:lightcone}
\begin{figure}
    \includegraphics[width=0.47\textwidth]{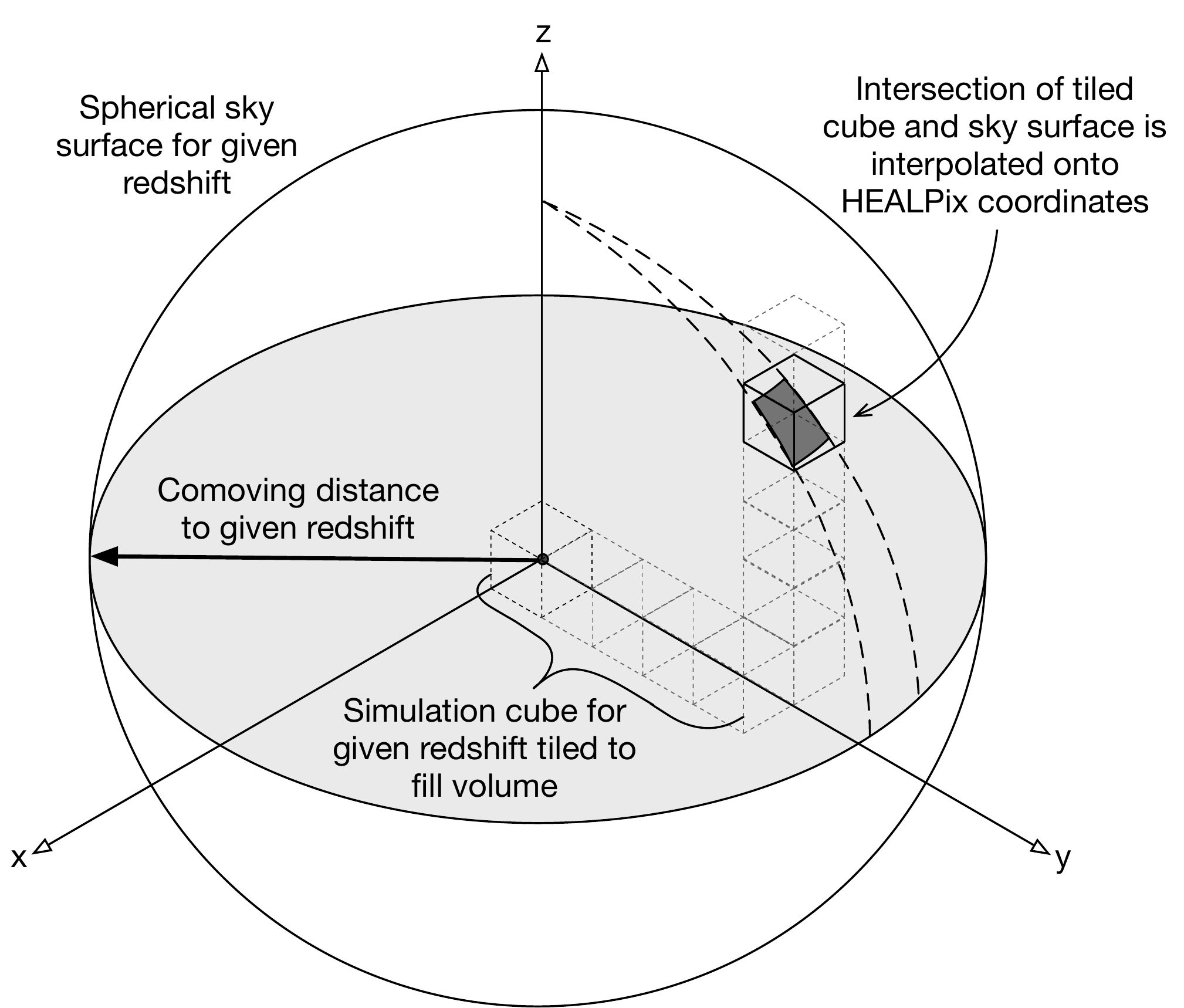}
    \caption{Schematic diagram illustrating the tiling of simulation cubes with periodic boundary conditions to produce a full-sky lightcone. The diagram shows the process for a single redshift. The process is repeated for every observed redshift using an evolved simulation cube for each redshift and the appropriate comoving distance diameter for each redshift. The (x, y, z) coordinates in the diagram represent both the comoving coordinates of the simulation cubes and the Cartesian coordinates of the HEALPix sphere.}
    \label{fig:tiling_scheme}
\end{figure}

The observable sky at a particular redshift is a sphere in comoving space.  No existing 21~cm simulation spans a volume large enough to contain the whole observable sky at the redshifts of interest for reionization. The typical size of a large 21~cm simulation box presently is $\sim$2 Gpc while a comoving distance to redshift 8.5 ($\nu_{obs}\approx150$ MHz) is $\sim$9.3 Gpc.  To create a full-sky 21~cm model, we exploit the periodic boundary condition of the 21~cm simulations and effectively tile the simulation volume in comoving space to obtain a sufficiently large volume before gridding on to the sky sphere.  The process is repeat at each redshift of interest to create maps for the full lightcone.   Figure~\ref{fig:tiling_scheme} shows a schematic diagram of our tiling and gridding process.

We start with a suite of theoretical 21~cm cubes from the semi-analytic simulations of \citet{2013ApJ...767...68M}. The cubes span the redshift range $9.3>z>6.2$, with $\Delta z=0.1$ steps, representing the universe from $\sim30\%$ to $96\%$ ionised. The simulation volume is $1\,\mathrm{Gpc}^3$ in a $512^3$ pixel box with periodic boundary. We linearly interpolate the simulated 21~cm cubes across redshift to produce new cubes that more-closely match the redshifts observed by HERA in steps of 80 kHz spectral channels between $\sim139-195$ MHz. The interpolation step is not required to construct the lightcone if simulation cubes matching the redshift of interests are available. For each of the redshifts, we tile the interpolated cube with itself in three dimensions to construct an arbitrarily large simulated volume for that redshift. Then, we draw an observable sky at that redshift as a sphere of radius equal to the comoving distance of that redshift from a fixed origin inside the volume and interpolate the nearest neighbouring pixel from the cube to the corresponding HEALPix pixel location on the sphere. Before tiling and gridding, we degrade each simulation cube to $128^3$ box to reduced computing time while retaining all size scales that can be probed by HERA. We use HEALPix pixel area that is $\sim10$ times smaller than the resolution of the simulations (NSIDE=4096) to avoid sampling artefacts. This process is repeated for every observed redshift to produce a suit of full-sky maps that accurately represent the redshifted 21~cm lightcone model.

\begin{figure*}
    \includegraphics[width=1.0\textwidth]{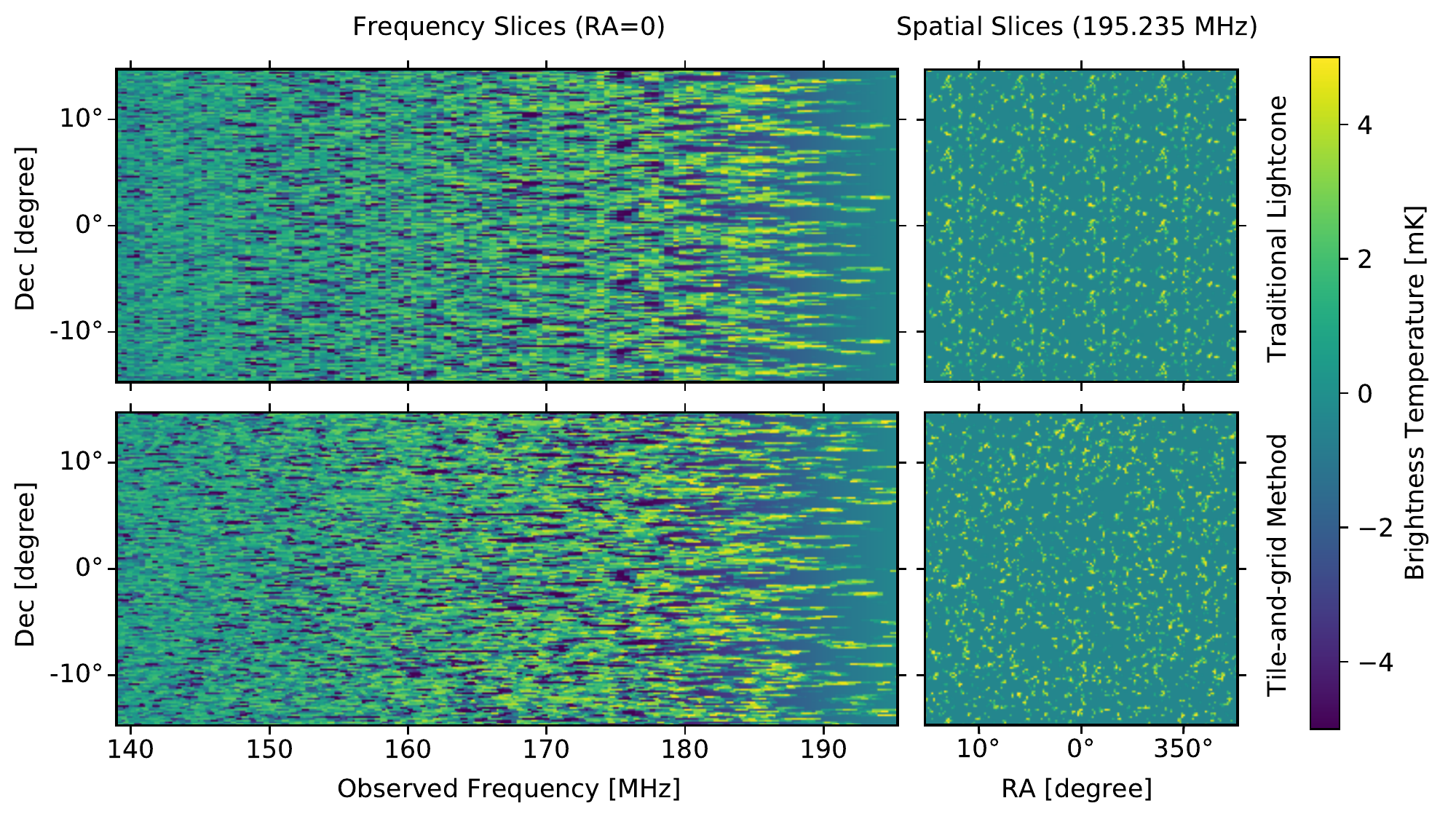}
    \caption{ Comparison of flat-field lightcone tiling with our spherical surface projection.  Top panel shows the flat-field lightcone and the bottom panel shows the spherical surface lightcone.  Even over the relatively small area plotted, repetition of structure is visible in the flat-field lightcone, whereas the spherical surface method results in a random appearance while still preserving the one-point statistics of the original simulation cubes.  The flat-field lightcone cube is interpolated from comoving (x, y, z) directly to celestial image coordinates. The two images are identical along Dec = 0 degree row.}
    \label{fig:tiling_compare}
\end{figure*}

Compared to flat-field approximations for tiling and gridding that do not take into account the spherical surface of the sky, our method is equivalent to slicing a simulation cube from different angles and rotational axes and mapping the slices onto a sphere at different locations. Thus, even within the $\sim9$~degree HERA fields, we see significantly reduced repetition of spatial structure in the resulted maps in comparison to the flat-field approximation.  Figure~\ref{fig:tiling_compare} illustrate this point, where we use the method from \citet{2014MNRAS.439.1615Z} to produce a standard flat-field lightcone cube and compare it with a lightcone cube produced by gridding onto spherical surfaces.   The full-sky HEALPix maps produced with spherical surface gridding preserve one-point statistics of the original simulations, showing changes less than 0.001$\%$ of the original values for our simulation sets.

\section{Kurtosis Uncertainty Propagation}
\label{apd:noise}
The method of uncertainty propagation for one-point statistics is first described in \citet{Watkinson:2014jv}. Here, we summarise and expand on their work, deriving uncertainty propagation for the kurtosis in addition to variance and skewness.

To recap, for a 21~cm intensity map with pixel value $x_i$, mean $\overline{x}$ and $N_{pix}$ pixels, the p-th central moment of the map is defined as, 
\begin{equation}
    m_p = \frac{1}{N_{pix}}\sum_{i=0}^{N_{pix}} (x_i - \overline{x})^p.
\end{equation}
The variance, skewness and kurtosis are standardisations of 2nd, 3rd and 4th central moments defined as follows.
\begin{align}
    \text{variance:\quad} &S_2 = m_2, \\
    \text{skewness:\quad} &S_3 = \frac{m_3}{(m_2)^{3/2}}, \\
    \text{kurtosis:\quad} &S_4 = \frac{m_4}{(m_2)^2} - 3. 
\end{align}
If every pixel $x_i$ consists of only an independent signal with no noise contribution, we can simply substitute $x_i=\delta T_i$ and $\overline{x}=\overline{\delta T}$ to compute the ``true'' moments and one-point statistics of the map. 

Adding noise $n_i$ with standard deviation $\sigma_i$ to the signal, each pixel now consists of the signal plus the noise, $x_i = \delta T_i + n_i$, and the noise will bias the moment measurements. An unbiased estimator for the p-th moments ($\hat{m}_p$) can be estimated by averaging the moment equations over noise realisation. Assuming that the noise is Gaussian and independent in each pixel, the averaged noise terms can be rewritten as functions of standard deviation of the noise, using Gaussian moment identities derivable from the following formula,
\begin{equation}\label{eq:gauss_id}
    \langle n_i^l \rangle = 
    \begin{cases}
    (1)(3)(5)\dotsm(l-1)\sigma_i^l & \text{if } l \text{ is even}\\
    0  & \text{if } l \text{ is odd}\\
    \end{cases}
\end{equation}
where the angle bracket designates an average. Table~\ref{tab:noise} in this work provide additional identities necessary for the derivation of the kurtosis uncertainty in addition to Equation~\ref{eq:gauss_id} and the identities given in the Table A1 in \citet{Watkinson:2014jv}.

Using these identities, the estimator variance and covariance of the unbiased estimator of the moments can be derived,
\begin{gather}
    V_{\hat{m_p}} = \langle\hat{m_p}\hat{m_p}^{\dagger}\rangle 
        - \langle\hat{m_p}\rangle^2, \label{eq:estimator_variance}\\
    C_{\hat{m}_p\hat{m}_q} = \langle\hat{m}_p\hat{m}_q\rangle 
        - \langle\hat{m}_p\rangle\langle\hat{m}_q\rangle,
\end{gather}
and propagate to skewness and kurtosis with Taylor expansion,
\begin{equation}\label{eq:taylor}
\begin{split}
    V_{f(X,Y)} \approx 
    \left(\frac{\partial f}{\partial X}\right)^2 
    V_{X}
    + \left(\frac{\partial f}{\partial Y}\right)^2
    V_{Y} \\
    + 2 \left(\frac{\partial f}{\partial X}\right) 
    \left(\frac{\partial f}{\partial Y}\right)
    C_{X Y}.
\end{split}
\end{equation}
Here, $f(X,Y)$ is a function of two non-independent variables $X$ and $Y$. In other word, $X=\hat{m_2}$ and $Y=\hat{m_3}$ for skewness, and $X=\hat{m_2}$ and $Y=\hat{m_4}$
for kurtosis. The uncertainty for each statistic is then just the square root of the estimator variance.

Equations~\ref{eq:old1} to \ref{eq:old2} summarise results from \citet{Watkinson:2014jv}. For this work, $\sigma_i$ is assumed to be equal to $\sigma_n$ in Equation~\ref{eq:noise2} in the main text for all pixels. We also simply use $N$ in the main text instead of $N_{pix}$ to emphasise that an individual pixel in an image from an observation may not represent an independent sample.
\begin{gather}
    \label{eq:old1}
    \hat{m}_2 = \frac{1}{N_{pix}} \sum_{i=0}^{N_{pix}}
        (x_i - \overline{x})^2 
        - \sigma_n^2, \\
    \hat{m}_3 = \frac{1}{N_{pix}} \sum_{i=0}^{N_{pix}}
        (x_i - \overline{x})^3 = m_3 = \frac{1}{N_{pix}} \sum_{i=0}^{N_{pix}}
        (\delta T_i - \overline{\delta T})^3, \\
    C_{\hat{m}_2\hat{m}_3} = \frac{6}{N_{pix}}m_3\sigma_n^2, \\
    V_{\hat{m}_2} = V_{\hat{S_2}} = \frac{2}{N_{pix}}
        (2 m_2 \sigma_n^2 + \sigma_n^4), \\
    V_{\hat{m}_3} = \frac{3}{N_{pix}}
        (3 m_4 \sigma_n^2 + 12 m_2 \sigma_n^4 
        + 5 \sigma_n^6), \\
    V_{\hat{S_3}} \approx \frac{1}{(m_2)^3}V_{\hat{m}_3}
        + \frac{9}{4}\frac{(m_3)^2}{(m_2)^5}V_{\hat{m}_2}
        - 3\frac{m_3}{(m_2)^4}C_{\hat{m}_2\hat{m}_3}. 
    \label{eq:old2}
\end{gather}

We follow the procedure in \citet{Watkinson:2014jv} and are able to confirm their results. In addition, we derive the estimator variance for kurtosis as follows. 

First the 4th moment with noise bias is constructed.
\begin{align}
    m_4^{biased} 
        &= \frac{1}{N_{pix}} \sum_{i=0}^{N_{pix}}
            (x_i - \overline{x})^4 \nonumber\\
        &= \frac{1}{N_{pix}} \sum_{i=0}^{N_{pix}}
            [(\delta T_i - \overline{\delta T}) 
            + n_i]^4.
\end{align}
Then, we expand the equation and average over the noise.
\begin{align}
    \langle m_4^{biased} \rangle
        &= \frac{1}{N_{pix}} \sum_{i=0}^{N_{pix}}
            [(\delta T_i - \overline{\delta T})^4 +
            4(\delta T_i - \overline{\delta T})^3 
            \langle n_i \rangle \nonumber\\
        &\quad+ 6(\delta T_i - \overline{\delta T})^2 
            \langle n_i^2 \rangle + 
            4(\delta T_i - \overline{\delta T}) 
            \langle n_i^3 \rangle + 
            \langle n_i^4 \rangle] \nonumber\\
        &= \frac{1}{N_{pix}} \sum_{i=0}^{N_{pix}}
            (\delta T_i - \overline{\delta T})^4 \nonumber\\
        &\quad+ 6 \frac{1}{N_{pix}} \sum_{i=0}^{N_{pix}}
            (\delta T_i - \overline{\delta T})^2 
            \sigma_{i}^2 + 3 \sigma_{i}^4 \nonumber\\
        &= \frac{1}{N_{pix}} \sum_{i=0}^{N_{pix}}
            (\delta T_i - \overline{\delta T})^4 + 6 m_2 \sigma_n^2 + 3 \sigma_n^4.\label{eq:biased_m4}
\end{align}
The first term in Equation~\ref{eq:biased_m4} is simply the 4th moment while other terms arise from the added noise. This implies that an unbiased estimator of the 4th moment is,
\begin{align}
    \hat{m}_4 
        &= \frac{1}{N_{pix}} \sum_{i=0}^{N_{pix}}
            (x_i - \overline{x})^4 
            - 6 m_2 \sigma_n^2 
            - 3 \sigma_n^4 \nonumber\\
        &= \frac{1}{N_{pix}} \sum_{i=0}^{N_{pix}}
            (x_i - \overline{x})^4 
            - \frac{3}{2}N_{pix}V_{\hat{m_2}}.
    \label{eq:unbiased_m4}
\end{align}

Next we derive the estimator variance of the 4th moment. We substitute $\mu_i=\delta T_i - \overline{\delta T}$ and $\kappa=3N_{pix}V_{\hat{m_2}}/2$ to simplify Equation~\ref{eq:unbiased_m4} before plugging into Equation~\ref{eq:estimator_variance} to obtain,
\begin{equation}
    \begin{split}
        V_{\hat{m}_4} = \biggl\langle \frac{1}{N_{pix}^2}
            \sum_{i=0}^{N_{pix}}\sum_{j=0}^{N_{pix}}
            \lbrack(\mu_i+n_i)^4-\kappa\rbrack \\
        \times \lbrack(\mu_j+n_j)^4-\kappa\rbrack
            \biggr\rangle - (m_4)^2,
    \end{split}
\end{equation}
where the second term is reduced to the square of the unbiased 4th moment.

Expanding this expression and moving the noise averaging brackets inside the summation gives,
\begin{align}
    V_{\hat{m}_4} = &\frac{1}{N_{pix}^2}
        \sum_{i=0}^{N_{pix}}\sum_{j=0}^{N_{pix}}
        \Big\lbrack\mu_i^4\mu_j^4 
        + 4\mu_i^4\mu_j^3\langle n_j \rangle \nonumber\\
        &+ 6\mu_i^4\mu_j^2\langle n_j^2 \rangle 
        + 4\mu_i^4\mu_j\langle n_j^3 \rangle 
        + \mu_i^4\langle n_j^4 \rangle 
        - \kappa\mu_i^4 \nonumber\\
        &+ 4\mu_i^3\mu_j^4\langle n_i \rangle
        + 16\mu_i^3\mu_j^3\langle n_i n_j \rangle 
        + 24\mu_i^3\mu_j^2\langle n_i n_j^2 \rangle\nonumber\\
        &+ 16\mu_i^3\mu_j\langle n_i n_j^3 \rangle 
        + 4\mu_i^3\langle n_i n_j^4 \rangle
        - 4\kappa\mu_i^3\langle n_i\rangle \nonumber\\
        &+ 6\mu_i^2\mu_j^4\langle n_i^2 \rangle 
        + 24\mu_i^2\mu_j^3\langle n_i^2n_j \rangle 
        + 36\mu_i^2\mu_j^2\langle n_i^2n_j^2 \rangle\nonumber\\ 
        &+ 24\mu_i^2\mu_j\langle n_i^2n_j^3 \rangle 
        + 6\mu_i^2\langle n_i^2n_j^4 \rangle
        - 6\kappa\mu_i^2\langle n_i^2\rangle \nonumber\\
        &+ 4\mu_i\mu_j^4\langle n_i^3 \rangle 
        + 16\mu_i\mu_j^3\langle n_i^3n_j \rangle 
        + 24\mu_i\mu_j^2\langle n_i^3n_j^2 \rangle\nonumber\\ 
        &+ 16\mu_i\mu_j\langle n_i^3n_j^3 \rangle 
        + 4\mu_i\langle n_i^3n_j^4 \rangle
        - 4\kappa\mu_i\langle n_i^3 \rangle \nonumber\\
        &+ \mu_j^4\langle n_i^4 \rangle 
        + 4\mu_j^3\langle n_i^4n_j \rangle 
        + 6\mu_j^2\langle n_i^4n_j^2 \rangle\nonumber\\ 
        &+ 4\mu_j\langle n_i^4n_j^3 \rangle 
        + \langle n_i^4n_j^4 \rangle
        - \kappa\langle n_i^4 \rangle \nonumber\\
        &- \kappa\mu_j^4
        - 4\kappa\mu_j^3\langle n_j \rangle
        - 6\kappa\mu_j^2\langle n_j^2 \rangle
        - 4\kappa\mu_j\langle n_j^3 \rangle \nonumber\\
        &- \kappa\langle n_j^4 \rangle
        + \kappa^2 \Big\rbrack - (m_4)^2.
\end{align}
Applying Gaussian noise identities will reduce the expression to,
\begin{align}
    V_{\hat{m}_4} = &\frac{1}{N_{pix}^2}
        \sum_{i=0}^{N_{pix}}\sum_{j=0}^{N_{pix}}
        \Big\lbrack\mu_i^4\mu_j^4 
        + 6\mu_i^4\mu_j^2\sigma_j^2 
        + 3\mu_i^4\sigma_j^4 
        - \kappa\mu_i^4 \nonumber\\
        &+ 16\mu_i^3\mu_j^3\delta_{ij}\sigma_j^2 
        + 48\mu_i^3\mu_j\delta_{ij}\sigma_i^4
        + 6\mu_i^2\mu_j^4\sigma_i^2 \nonumber\\ 
        &+ 36\mu_i^2\mu_j^2(1+2\delta_{ij})\sigma_i^2\sigma_j^2 
        + 6\mu_i^2(3+12\delta_{ij})\sigma_i^2\sigma_j^4
        \nonumber\\
        &- 6\kappa\mu_i^2\sigma_i^2 
        + 48\mu_i\mu_j^3\delta_{ij}\sigma_j^4
        + 240\mu_i\mu_j\delta_{ij}\sigma_i^6 \nonumber\\ 
        &+ 3\mu_j^4\sigma_i^4 
        + 6\mu_j^2(3+12\delta_{ij})\sigma_i^4\sigma_j^2 
        + (9+96\delta_{ij})\sigma_i^4\sigma_j^4 \nonumber\\
        &- 3\kappa\sigma_i^4
        - \kappa\mu_j^4
        - 6\kappa\mu_j^2\sigma_j^2
        - 3\kappa\sigma_j^4 
        + \kappa^2 \Big\rbrack \nonumber \\
        &- (m_4)^2.
\end{align}
Doing the summation to perform index conversion via $\delta_{ij}$, substituting all $\frac{1}{N_{pix}}\sum_{i=0}^{N_{pix}}\mu_i^k$ terms with the unbiased p-th moments $m_p$ and $\sigma_i$ with $\sigma_n$, re-substituting $\kappa = 3N_{pix}V_{\hat{m_2}}/2 = 6 m_2 \sigma_n^2 + 3 \sigma_n^4$ back to the expression, and cancelling out many terms will yield the estimator variance of the 4th moment,
\begin{equation}
\begin{split}
    V_{\hat{m}_4} = \frac{8}{N_{pix}}
    (2 m_6 \sigma_n^2 + 21 m_4 \sigma_n^4 \\
    + 48 m_2 \sigma_n^6 + 12 \sigma_n^8).
    \label{eq:var_m4}
\end{split}
\end{equation}

The estimator covariance between 2nd and 4th moment can be found in a similar manner, resulting in,
\begin{equation}
    C_{\hat{m}_2\hat{m}_4} = \frac{4}{N_{pix}}
        (2 m_4 \sigma_n^2 + 9 m_2 \sigma_n^4
        + 3 \sigma_n^6).
    \label{eq:covar_m2_m4}
\end{equation}

Finally, we can propagate Equation~\ref{eq:var_m4} and \ref{eq:covar_m2_m4} using Equation~\ref{eq:taylor} to obtain the estimator variance for kurtosis,
\begin{equation}
    V_{\hat{S_4}} = \frac{1}{(m_2)^4} V_{\hat{m}_4}
    + 4 \frac{(m_4)^2}{(m_2)^6} V_{\hat{m}_2} - 4 \frac{m_4}{(m_2)^5} C_{\hat{m}_2\hat{m}_4}.
\end{equation}

\begin{table}
\caption{Additional Gaussian noise identities for derivation of estimator variance of kurtosis. Please see Table A1 in \citet{Watkinson:2014jv} for more identities.}
\label{tab:noise}
\begin{tabular}{l l}
    \hline
    $\begin{array}{l l l l}
        \langle n_i n_j^4 \rangle 
            && \langle n_i^5 \rangle=0 & (i=j) \\
            && \langle n_i \rangle\langle n_j^4 \rangle=0 & (i \neq j)
    \end{array}$ & $\bigg\rbrace0$ \\
    $\begin{array}{l l l l}
        \langle n_i^2 n_j^4 \rangle 
            && \langle n_i^6 \rangle=15\sigma_i^6 & (i=j) \\
            && \langle n_i^2 \rangle\langle n_j^4 \rangle=3\sigma_i^2\sigma_j^4 & (i \neq j)
    \end{array}$ & $\bigg\rbrace(3+12\delta_{ij})\sigma_i^2\sigma_j^4$ \\
    $\begin{array}{l l l l}
        \langle n_i^3 n_j^4 \rangle 
            && \langle n_i^7 \rangle=0 & (i=j) \\
            && \langle n_i^3 \rangle\langle n_j^4 \rangle=0 & (i \neq j)
    \end{array}$ & $\bigg\rbrace0$ \\
    $\begin{array}{l l l l}
        \langle n_i^4 n_j^4 \rangle 
            && \langle n_i^8 \rangle=105\sigma_i^8 & (i=j) \\
            && \langle n_i^4 \rangle\langle n_j^4 \rangle=9\sigma_i^4\sigma_j^4 & (i \neq j)
    \end{array}$ & $\bigg\rbrace(9+96\delta_{ij})\sigma_i^4\sigma_j^4$ \\
    \hline
\end{tabular}
\end{table}


\bsp	
\label{lastpage}
\end{document}